\documentclass[12pt]{article}
\usepackage{amsmath,amsfonts,amssymb,latexsym}
\def\references{
\vskip 0.2cm
\vskip \baselineskip
\addcontentsline{toc}{section}{References}
\phantom{?}\vspace{-2\baselineskip}}

\def\acknowledgements{
\vskip 0.2cm
\vskip \baselineskip
\addcontentsline{toc}{section}{Acknowledgements}
\phantom{?}\vspace{-2\baselineskip}
\section*{Acknowledgements}}

\makeatletter

\def\tableofcontents{%
  \section*{\contentsname}%
\vskip 0.2cm
  \@starttoc{toc}\vskip 1.2cm}

\renewcommand\l@section[2]{%
  \ifnum \c@tocdepth >\z@
    \addpenalty\@secpenalty
    \addvspace{1.0em \@plus\p@}%
    \setlength\@tempdima{1.5em}%
    \begingroup
      \parindent \z@ \rightskip \@pnumwidth
      \parfillskip -\@pnumwidth
      \leavevmode {\bfseries
      \advance\leftskip\@tempdima
      #1}\nobreak\ 
      \leaders\hbox{$\m@th\mkern \@dotsep mu\hbox{.}\mkern \@dotsep mu$}
     \hfil \nobreak\hb@xt@\@pnumwidth{\hss\bfseries #2}\par
    \endgroup
  \fi}

\renewcommand{\section}{\@startsection
  {section}
  {1}
  {0mm}
  {-\baselineskip}
  {0.75\baselineskip}
  {\bfseries\Large}
}%

\renewcommand{\subsection}{\@startsection
  {subsection}%
  {2}%
  {0mm}%
  {-\baselineskip}%
  {0.5\baselineskip}%
  {\bfseries\large}%
}%

\renewcommand{\subsubsection}{\@startsection
  {subsubsection}%
  {3}%
  {0mm}%
  {-\baselineskip}%
  {0.5\baselineskip}
  {\bfseries\normalsize}%
}%

\makeatother
\numberwithin{equation}{section}
\long\def\symbolfootnote[#1]#2{\begingroup%
\def\thefootnote{\fnsymbol{footnote}}\footnote[#1]{#2}\endgroup}

\newcommand{\be}{\begin{equation}}
\newcommand{\ee}{\end{equation}}
\newcommand{\ba}{\begin{aligned}}
\newcommand{\ea}{\end{aligned}}
\renewcommand{\a}{\alpha}
\renewcommand{\b}{\beta}
\newcommand{\g}{\gamma}
\renewcommand{\d}{\delta}
\newcommand{\e}{\epsilon}

\renewcommand{\t}{\theta}
\renewcommand{\l}{\lambda}
\newcommand{\m}{\mu}
\newcommand{\n}{\nu}
\renewcommand{\r}{\rho}
\newcommand{\s}{\sigma}
\renewcommand{\L}{\Lambda}
\renewcommand{\O}{\Omega}

\newcommand{\G}{\Gamma}

\newcommand{\p}{\partial}
\newcommand{\trans}{^{\hspace{0.1mm}\ensuremath{\mathsf{T}}}}
\newcommand{\sm}[2]{$\mathcal{M}^{\,\scriptscriptstyle(#1|#2)}$}
\def\zero{\,^{\scriptscriptstyle 0}}
\def\one{\!^{\scriptscriptstyle 1}}
\def\two{\,^{\scriptscriptstyle 2}}
\newcommand{\vare}{\varepsilon}
\newcommand{\tr}{\mathrm{Tr\,}}
\newcommand{\De}{\mathrm{D}}

\long\def\symbolfootnote[#1]#2{\begingroup%
\def\thefootnote{\fnsymbol{footnote}}\footnote[#1]{#2}\endgroup}

\begin{document}
\thispagestyle{empty}
\rightline{\tt hep-th/0607243}\vskip -0.3mm
\rightline{\tt DISTA-UPO-06}\vskip -0.3mm
\rightline{\tt DFTT-18/2006}\vskip -0.3mm
\rightline{\tt July, 2006}
\vskip 0.8cm
\hrule
\vskip 1cm\noindent
{\hskip 0.4 cm\bfseries\LARGE Flux Vacua and Supermanifolds}
\vskip 1.2cm\noindent
{\hskip 0.4 cm\bfseries P. A. Grassi}$^{\,1,2,3,5,}\symbolfootnote[2]{\tt pgrassi@cern.ch}$ {\bfseries and M. Marescotti}$^{\,4,5,}$\symbolfootnote[3]{\tt marescot@to.infn.it}
\vskip .3 truecm
\small\noindent\small
\hskip 0.4 cm$^1$\,CERN, Theory Unit, CH-1211 Geneve, 23, Switzerland
\vskip .03cm\noindent
\hskip 0.4 cm$^2$\,Centro Studi e Ricerche E. Fermi,
Compendio Viminale, I-00184, Roma, Italy
\vskip .03cm\noindent
\hskip 0.4 cm$^3$\,DISTA, Universit\`a del Piemonte Orientale,
Via Bellini 25/{\scriptsize G}, I-15100, Alessandria, Italy
\vskip .03cm\noindent
\hskip 0.4 cm$^4$\,Dipartimento di Fisica Teorica, Universit\`a di Torino, Via Giuria 1, I-10125, Torino, Italy
\vskip .03cm\noindent
\hskip 0.4 cm$^5$\,INFN, sezione di Torino, Via P. Giuria 1, I-10125, Torino, Italy\\
\vskip 0.2cm
\normalsize
\hrule\vskip 1.3cm
\subsubsection*{Abstract}
As been recently pointed out, physically relevant models derived from 
string theory require the presence of non-vanishing form fluxes besides 
the usual geometrical constraints. In the case of NS-NS fluxes, the 
Generalized Complex Geometry encodes these informations in a beautiful 
geometrical structure. On the other hand, the R-R fluxes call for supergeometry 
as the underlying mathematical framework. In this context, we analyze the possibility of constructing interesting supermanifolds recasting the geometrical 
data and RR fluxes. To characterize these supermanifolds we have been 
guided by the fact topological strings on supermanifolds require the super-Ricci flatness of the target space. This can be achieved 
by adding to a given bosonic manifold enough anticommuting coordinates and 
new constraints on the bosonic sub-manifold. We study these constraints 
at the linear and non-linear level for a pure geometrical setting and in the presence of p-form field strengths. We find that certain spaces admit 
several super-extensions and we give a parameterization in a simple case 
of d bosonic coordinates and two fermionic coordinates. In addition, we 
comment on the role of the RR field in the construction of the super-metric. 
We give several examples based on supergroup manifolds and 
coset supermanifolds.
\newpage

{\section{Introduction}

Recently the pure spinor 
formulation of superstrings \cite{Berkovits:2000fe,Berkovits:2002zk} and the twistor string 
theory \cite{Witten:2003nn} have promoted  supermanifolds as a 
fundamental tool to  formulate the string theory and supersymmetric models. 
The main point 
is the existence of background fields of extended supergravity theory 
(and therefore of 10 dimensional superstrings) which are the $p$-forms 
of the fermionic sector of superstrings. 
Those fields can be coupled to worldsheet sigma models fields by considering 
the target space as a supermanifold. 
Indeed, as shown in \cite{Berkovits:1999im,Berkovits:1999zq}
by adding to the bosonic space 
some anticommuting coordinates  $\theta$'s, one can easily encode the informations 
regarding the geometry and the $p$-form fluxes into the supermetric of a supermanifold. 
In the present paper, we provide a preliminary analysis of the 
construction of supervarieties on a given 
bosonic space with or without $p$-forms (these are usually 
described by a bispinor $F^{\mu\nu}$ since they couple 
to target space spinors). We have to mention the interesting 
results found in \cite{Rocek:2004bi,Rocek:2004ha,Zhou:2004su,Lindstrom:2005uh}
stimulated by the work \cite{Witten:2003nn,Aganagic:2004yh}. There the case of super-CY 
is analyzed and it is found that by limiting the number of fermions, the constraints 
on the bosonic submanifold of the supermanifold become very stringent. The super-CY spaces 
were introduced in string theory in paper \cite{Sethi:1994ch}. There the sigma models with supertarget spaces, their conformal invariance and the analysis of some topological rings were studied. In addition, 
we would like also to make a bridge with the recent successes of the Generalized Complex 
Geometry \cite{Hitchin:2004ut,Gualtieri:2003dx} to construct N=1 supersymmetric vacua of type 
II superstrings \cite{mich}. In that context, the flux of the NS-NS background contributes to the generalized geometry 
and space is no longer CY. In the same way for a super-CY its bosonic subspace does not need to be a CY.

The differential geometry of supermanifolds is a generalization of the usual 
geometry bosonic manifolds extended to anticommuting coordinates. There 
one can define supervector fields, superforms and the Cartan calculus. In 
addition, one can extend the usual Levi-Civita calculus to the 
fermionic counterparts defining super-Riemann, super-Ricci and 
super-Weyl tensor fields. A tensor has fermionic components and 
each component is a superfield. Due to the anticommuting coordinates, 
one defines also a metric on a supermanifold which can be 
generically given by 
\begin{equation}\label{introA}
ds^2 = G_{(mn)} dx^m \otimes 
dx^n + G_{m \mu} dx^m \otimes d\theta^\mu +  G_{[\mu \nu]} 
d\theta^\mu \otimes d\theta^\nu
\end{equation}
where $G_{(mn)}(x,\theta), G_{m\mu}(x,\theta), G_{[\mu \nu]}(x,\theta)$ are superfields and the tensor product $\otimes$ respects the parity of 
the differentials $dx^{m}, d\theta^{\mu}$. The superfields are polynomials 
of $\theta$'s and the coefficients of their expansions are ordinary bosonic 
and fermionic fields. For example the expansion of $G_{mn}(x,\theta) = 
G^{(0)}_{mn}(x) + \dots $ has a direct physical interpretation: $G^{(0)}_{mn} = g^{(body)}_{mn}(x)$ where the latter is the metric of the bosonic submanifold and the 
higher components are fixed by requiring the super-Ricci flatness. The physical interpretation of $G_{\mu m}$ and $G_{\mu\nu}$ is more interesting. Indeed, 
as suggested by the pure spinor string theory 
\cite{Berkovits:2000fe,Berkovits:2002zv,Grassi:2005av} 
the first components of $G_{\mu\nu}$ 
can be identified with a combination of RR field strengths (in the text, this 
point will be clarified further). The first component of $G_{\mu m}$ are fixed by consistency and, in  a suitable gauge, it coincides with a Dirac matrix. 

This 
is not the only way in which the RR fields determine the geometrical 
structure of the supermanifold.  We learnt from
recent analysis \cite{Klemm:2001yu,deBoer:2003dn,Ooguri:2003qp,Seiberg:2003yz} in superstrings and, consequently 
in supersymmetric field theory, that the superspace can be 
deformed by the existence of RR field strength $F^{\mu\nu}(x,\theta)$
\be\label{introB}
\{ \theta^{\mu}, \theta^{\nu} \} = F^{\mu\nu}\,.
\ee
This has been only verified for constant RR field strengths. 

Notice that a given bispinor $F^{\mu\nu}$ is decomposed 
into a symmetric $F^{(\mu\nu)}$ and antisymmetric part $F^{[\mu\nu]}$. Therefore, it is 
natural to identify the antisymmetric part with 
the fermion-fermion (FF) components $G_{[\mu\nu]}(x)$ 
of the supermetric (\ref{introA}) and the symmetric one with the 
non-vanishing r.h.s. of (\ref{introB}). This is very satisfactory since it 
provides a complete mapping between the superstring background 
fields and the deformations of the superspace: the non-commutativity and 
the non-flatness. Indeed, the presence of a superfield $G_{\mu\nu}(x,\theta)$ 
in the supermetric modifies the supercurvature of the superspace. 
So, finally we can summarize the situation in the following way: 
given the backgrounds $g_{mn}$ and the NSNS $b_{mn}$ they are responsible 
for the deformation of the bosonic metric and of the commutation 
relations between bosonic coordinates $[ x^{m}, x^{n}] = \theta^{[mn]}$ 
(where $\theta^{mn} = (b^{-1})^{mn}$). On the other side, the 
RR fields deforms the supermetric and the anticommutation relations. 

There is another important aspect to be consider. The anticommuting 
deformation of superspace is treatable whether the field strength $F^{\mu\nu}$ 
of the RR field is constant and, in four dimensions \cite{Ooguri:2003qp}, 
whether is self-dual. In that case, there is no back-reaction of the 
metric and it is a solution of the supergravity equations. In the case of 
antisymmetric RR fields, which deform the supermetric (\ref{introA}), we have 
to impose a new condition in order that the worldsheet 
sigma model is conformal and therefore treatable. 
We require that the supermanifold is super-Ricci flat 
\begin{equation}\label{introC}
R_{MN} =0\,.
\end{equation}
In this way, we can still use some of the conventional technique 
of conformal field theory and of topological strings to study such models. 
The main point is that even if there is back-reaction, this is  
under control since the total superspace is super-Ricci flat 
\cite{Berkovits:1999im,Berkovits:1999zq,Bershadsky:1999hk,Seki:2006cj,Kagan:2005wt}. 

We would like to remind the reader that the super-Ricci flatness is 
the condition which guarantees the absence of anomalies in the 
case of B-model topological strings on supermanifolds \cite{Hori:2003ic}. This is 
equivalent to the well-known condition for conventional B-model 
on bosonic target spaces. The latter have to be Calabi-Yau which 
are Ricci-flat K\"ahler spaces. A main difference is that the 
CY must be a 3-fold in order to compensate the ghost anomaly. 
In the case of supermanifold this is given by a virtual dimension 
which is essentially the difference between the bosonic and fermionic dimensions 
\cite{Ricci:2005cp}. In the A-model, the CY conditions is required for open 
string in presence of D-branes (see \cite{Hori:2003ic} for a complete discussion).\footnote{We would 
like also to recall the work \cite{Kapustin:2004gv,Kapustin:2004gv,Zucchini:2004ta,Lindstrom:2004iw,Zabzine:2005qf} 
where the construction of worldsheet sigma models on generalized CY is developed. 
}

We start with
a metric on a given space, for example 
the round sphere metric for $S^{n}$, or Fubini-Study metric 
for ${CP}^{n}$. Then we construct the corresponding supermanifold 
by adding anticommuting coordinates and additional components 
of the metric (in some cases, the extension of the metric to a 
super-metric is not enough and some non-vanishing torsion components are necessary) 
and we require the super-Ricci flatness.  
As a consequence, we notice that the first coefficient of $G_{\m\n}$ must be different 
from zero. Namely, we need a non-vanishing component of the fermion-fermion part of the metric. 
However, in general for a given space, no invariant spinorial tensor exists to provide such component. 
Therefore, we argue that this should be identified with $p$-form fluxes needed to sustain the solution 
of the supergravity. As a matter of fact, super-Ricci flatness does not seems to relate the value of the flux with the 
geometry of the bosonic submanifold. In fact, we will see in the forthcoming sections, the value of the flux is fixed by a 
suitable normalization which is implemented by a gauge condition on the supermetric. So, we conclude that 
super Ricci flatness turns out to be a necessary ad useful condition, but it does not seem to be sufficient 
to characterize the vacuum of the theory. 

We have to divide the analysis in two parts: the 
first one is the analysis of linearized equations and 
the second one concerns the analysis of the non-linear theory. 
It turns out that 
for the linear equation one gets very strong constraints on the 
bosonic metric such as the scalar curvature must be zero (see also 
\cite{Rocek:2004bi,Rocek:2004ha,Lindstrom:2005uh}). This is 
compatible with the flat or Ricci flat space, but certainly is not 
true for a sphere or any Einstein manifold. Therefore, it is unavoidable to tackle 
the non-linear analysis to see whether these constraints can be weakened. Let us explain briefly the lines of analysis. 
The Ricci-flatness condition (\ref{introC}) can be analyzed by expanding 
the super-metric in components. Since we are interested only  in the 
vacuum solutions, we decompose the superfields of the super-metric into 
bosonic components by setting to zero all fermionic ones. This 
implies a set of equations for each single component. Among them, 
a subset of these equations can be easily solved; indeed, 
the equations for the components to the order $n$ 
are solved in terms of some components 
at the $n+1$ order. 
However,  the equations for highest-components cannot be solved algebraically and 
in fact, by plugging  
the solutions of the lower orders in these equations, one finds high-derivative 
equations for the lowest components. 

To show that our analysis is not purely academic, we give some examples of supermanifolds 
modeled on some bosonic submanifold. Of course, the most famous example is $PSU(2,2|4)$ 
(from AdS/CFT correspondence) and $SL(4|4)$ (appearing in twistor string theory). Moreover, 
we give examples based on $S^{n}, V^{(p,q)}$ and $\mathbb{CP}^{n}$. As a by-product 
we construct a super-Hopf fibration connecting the different types of supercoset manifolds. 
 One interesting example is the construction of a supermanifold with $T^{(1,1)}$ as a bosonic 
 submanifold. In addition, we propose a construction which can be applied to any Einstein 
 spaces, and in particular to Sasaki-Einstein space and K\"ahler-Einstein. At the moment 
 we have not explored the role of Killing spinors, Killing vectors and the role of supersymmetry 
 in the construction. We only argued that if the supermanifold is obtained as a single coset space 
 and not a direct product of coset space, it has more chances to have an enhanced supersymmetry. 
 The relation between supersymmetry and the construction of super-Ricci flat supermanifold 
 will be explored in forthcoming papers. More important, it is not yet 
 established the relation between the solution of supergravity and the super-Ricci flatness.

The paper is organized as follows: in sec. \ref{sec1} we recall some 
basic fact about supergeometry. . In sec. \ref{sec2}, we present a general strategy, 
we study the linearized version of the constraints and, in the case of $d$ bosonic coordinates and 2 fermionic 
directions, we analyze completely the non-linear system. 
In sec. \ref{sec3}, we study the construction of supercosets modeled 
on bosonic cosets. In sec.  \ref{sec4}, we add the fluxes (RR field strengths). 
Appendices A and B contain some related material. 


\section{Elements of supergeometry}
\label{sec1}

We adopt the definition of supermanifold given in \cite{DeWitt:1992cy} based on a superalgebra with 
the coarse topology. However, this is not the only way and there is an equivalent formulation
based on sheaves and flag manifolds \cite{Bartocci:1988,Manin:1988ds}. The 
equivalence turns out to be useful for studying topological 
strings on supermanifolds \cite{math-dista}. 

\subsection{Supermanifolds}

A supermanifold \sm{m}{n} is called Riemannian if it is endowed with a metric supertensor which is 
a real non-singular commuting tensor field $G_{MN}$ satisfying the following graded-symmetry condition:
\be\label{graded-symmetry}
G_{MN}=(-1)^{MN}G_{NM}\,.
\ee
The meaning of \eqref{graded-symmetry} becomes more evident if one distinguishes between bosonic and fermionic components 
and rewrites the metric supertensor as a block supermatrix $G\,$
\be
G_{MN}=\left(
\begin{array}{c|c}
g_{mn}&h_{m\n}\\[1mm]
\hline\\[-4mm]
h\trans_{\n m}&j_{\m\n}\\
\end{array}\right)
\ee
being $\,g$ an $m\times m$ real symmetric commuting matrix, $j$ an $n\times n$ imaginary antisymmetric commuting matrix and $h$ an $m\times n$ imaginary anticommuting one. Obviously the first component of the superfield of the even-even block $g$ 
is an ordinary metric for the body $\mathcal{M}^{\,\scriptscriptstyle(m)}$ of the supermanifold \sm{m}{n}. Body means 
the bosonic submanifold.

Note that the supermetric $G$ is preserved by the orthosymplectic supergroup $\text{OSp}(m|n)$ whose bosonic submanifold is the direct product $\text{SO}(m)\times\text{Sp}(n)$ of an orthogonal group encompassing the Lorentz transformations of the metric $g$ with a symplectic group mixing the odd directions.
Thus, 
we immediately see that the number $n$ of fermionic dimensions (in the paper this is also denoted by $d_F$) 
over a Riemannian supermanifold must be even, otherwise $\text{Sp}(n)$ would not be defined.

The vielbein formalism can be constructed in the same way: supervielbeins will carry a couple of superindices whose different parity combinations induce the matrix block form
\be
E^A_{~M}=\left(
\begin{array}{c|c}
E^{\,a}_{~m}&E^{\,a}_{~\m}\\[1mm]
\hline\\[-4mm]
E^{\,\a}_{~m}&E^{\,\a}_{~\m}\\
\end{array}\right)
\ee
with commuting diagonal blocks and anticommuting off-diagonal blocks. The correct formula to pass from supervielbeins to supermetrics is
\be\label{inverse}
G_{MN}=(-1)^{MA}\,E^A_{~M}\,\eta_{AB}\,E^B_{~N}
\ee
where $\eta$ denotes the flat superspace metric $\,\text{diag}(\,\overbrace{\,\mathbf{1},\dots,\mathbf{1}\,,-\mathbf{1},\dots,-\mathbf{1}\,}^{m}\,,\,
\overbrace{\,\s_2,\dots,\s_2\,}^{n}\,)$\,. But what is $\eta_{\mu\nu}$? A question 
that has been raised several times in the 
literature (see for example \cite{VanNieuwenhuizen:1981ae} for a 
review and some comments), and it turns out that the flat 
super-metric in the fermionic directions $\eta_{\mu\nu}$ cannot be always 
defined for any models. For example, in the case of 10d supergravity IIB there 
is no constant tensor which can be used to raise and lower the 
spinorial indices and to contact to Weyl indices. 
However, it is also known that a typical bispinor 
emerges in the quantization of superstrings in the fermionic sector 
of the theory and these states are known as the RR fields. Indeed, 
in the case of constant RR fields, one can identify the 
fermionic components of the flat supermetric with constant RR field 
strength. This will be discuss in sec. \ref{sec4}. 

The inverse of a supermetric $G^{MN}$ is computed 
by inverting the supermatrix $G_{MN}$. In order 
that the inverse exists it must be that $\det(G_{mn})$ and $\det(G_{\mu\nu})$ 
do not vanish. 


\subsection{K\"ahler supermetrics}
A very important subclass of supermetrics consists of so-called K\"ahler supermetrics, which are a straightforward generalization of the usual K\"ahler metrics living on ordinary manifolds. As one can expect, the components $G_{M\bar{N}}$ of such a supermetric are obtained applying superderivatives - denoted by ``\,,\,'' - to a superpotential ${\cal K}$\,:
\be
G_{M\bar{N}}={\cal K}_{,\,M\bar{N}}\,.
\ee
As a consequence the body of a K\"ahler supermanifold with a supermetric $G_{M\bar{N}}$ originated by a superpotential ${\cal K}$ 
is a K\"ahler manifold endowed with the metric $g_{m\bar{n}}$ obtained by applying two derivatives to the bosonic part of ${\cal K}$ 
according to $g_{m\bar{n}}^{\scriptscriptstyle\,(body)}=\p_m\p_{\bar n}\,{\cal K}_{|\theta =0}\,$.
 

\subsection{Superconnections and covariant superderivative}
The idea of connection and covariant derivative have their relative supergeometric counterparts. 
Indicating superconnections with $\G$, one writes the defining rules for covariant superderivatives - denoted by ``\,;\,'' - as follows:
\begin{subequations}
\be
\Phi_{;M}=\Phi_{,M}\,,
\ee
\be
X^M_{~~;\,N}=X^M_{~~,\,N}+(-1)^{P(M+1)}X^P\,\G^{\,M}_{~PN}\,,
\ee
\be
\Phi_{M;\,N}=\Phi_{M,\,N}-\Phi_{P}\,\G^{\,P}_{~MN}\,.
\ee
\end{subequations}
A superconnection over a supermanifold \sm{m}{n} can be chosen in many ways. Usually this freedom is restricted by demanding that the superconnection be metric-compatible, i.e. that the covariant superderivative of the supermetric vanishes, 
and that the coefficients $\G^{\,M}_{~NP}$ be graded-symmetric in their lower indices. These two requirements uniquely determine the superconnection to be given by the super Christoffel symbols
\be\label{super-christoffel}
\left\{^{\phantom{?}M}_{N\!\phantom{?}\!P}\right\}=\,\frac{1}{2}\,
(-1)^Q\,
G^{\,MQ}\,\,[\,G_{QN,P}+(-1)^{NP}G_{QP,N}-( 
-1)^{Q(N+P)}\,G_{NP,\hspace{0.4mm}Q}\,]\,.
\ee
They obviously coincide with the usual Christoffel symbols over the body $\mathcal{M}^{\,\scriptscriptstyle(m)}$. The same conditions can also be rephrased in terms of the supervielbein $E^A$ and the spin-superconnection $\O^A_{~B}$\,:
\be\label{defspinconn}
dE^A+\O^A_{~B}\wedge E^B=0\,.
\ee
The superspin connection $\Omega^A_B$ is a 1-form. 


\subsection{Riemann and Ricci supercurvatures}

Taking care of signs due to different parities, Riemann and Ricci curvatures can be easily generalized to supermanifolds by the formulae
\be
\ba
R^{\,M}_{~NPQ}=&-\G^{\,M}_{~NP,\,Q}+(-1)^{P(N+R)}\,\G^{\,M}_{~RP}\,\G^{\,R}_{~NQ}\\&+(-1)^{PQ}\,\G^{\,M}_{~NQ,\,P}-(-1)^{Q(N+P+R)}\,\G^{\,M}_{~RQ}\,\G^{\,R}_{~NP}
\ea
\ee
and
\be\label{Ricci-tensor}
\ba
R_{MN}&\ba=(-1)^{P(M+1)}\,R^{\,P}_{~MPN}\ea\\&\ba=(-1)^{P(M+1)}\Big[\!&-\G^{\,P}_{~MP,\,N}+(-1)^{P(M+Q)}\,\G^{\,P}_{~QP}\,\G^{\,Q}_{~MN}\\&+(-1)^{NP}\,\G^{\,P}_{~MN,\,P}-(-1)^{N(M+P+Q)}\,\G^{\,P}_{~QN}\,\G^{\,Q}_{~MP}\Big]\,.\ea
\ea
\ee
Alternatively the Riemann supertensor can be defined as a differential superform by using the spin-superconnection $\O^A_{~B}$:
\be\label{defriemann}
d\O^A_{~B}+\O^A_{~C}\wedge\O^{\,C}_{~B}=R^A_{~B}\,.
\ee
If the superconnection does verify metric-compatibility and graded-symmetry conditions, Riemann and Ricci supertensors respectively satisfy the algebraic properties
\be
R_{MNPQ}=-\,(-1)^{MN}R_{NMPQ}=-\,(-1)^{PQ}R_{MNQP}=(-1)^{(M+N)(P+Q)}R_{PQMN}
\ee
and
\be
R_{\,MN}=(-1)^{MN}R_{\,NM}\,,
\ee
while Bianchi identities read
\be
R_{\,MNPQ;R}+(-1)^{P(Q+R)}R_{\,MNQR;P}+(-1)^{R(P+Q)}R_{\,MNRP;\,Q}=0\,,
\ee
\be
R_{\,MN;}^{\,\,\phantom{MN;}N}-\frac{1}{2}\,\,R_{\,;\,M}=0\,,
\ee
where $R=R_{MN}\,G^{NM}$ is the curvature super-scalar. 
In terms of supervielbeins, the 
vanishing of super-Ricci tensor can be written as 
\begin{equation}\label{addA}
E^{M}_{~A} R^{A}_{~B [MN\}} =0\,.
\end{equation}
The tensor 
$R^{A}_{~B M N} E^{M}_{~C} E^{N}_{~D} = R^{A}_{~BCD}$ is the 
curvature of the superspin connection (see 
\cite{Wess:1992cp}). $R=R_{MN}\,G^{NM}$ is the curvature super-scalar and 
there is a very compact formula for the Ricci curvature of K\"ahler supermanifolds 
\be
R_{M\bar{N}}=-\,(\,\ln \,\text{sdet}\, G)_{,\,M\bar{N}}\,.
\ee
which mimics the well-known formula of bosonic K\"ahler manifolds. 


\subsection{Supertorsion}
Retaining the metric compatibility and relaxing the graded-symmetry condition, in general superconnections depend on the supermetric and the supertorsion
\be
T^{\,M}_{~NP}=\G^{\,M}_{~NP}-(-1)^{NP}\,\G^{\,M}_{~PN}\,.
\ee
To be precise, a generic metric-compatible superconnection is the sum 
\be
\G^{\,M}_{~NP}=\left\{^{~\,M}_{\,N\,P\,}\right\}+K^{M}_{~NP}
\ee
of the Levi-Civita superconnection \eqref{super-christoffel} with so called supercontorsion $K$ defined by
\be
K^{M}_{~NP}=\frac{1}{2}\,\left(\,T^{\,M}_{~NP}-(-1)^{Q(N+1)}G^{MQ}G_{NR}\,T^{\,R}_{~QP}-(-1)^{Q(P+1)+NP}G^{MQ}G_{PR}\,T^{\,R}_{~QN}\,\right).
\ee
The corresponding Ricci supercurvature can be obtained adding a few terms to the expression valid with vanishing torsion:
\be
\ba
R_{MN}=\eqref{Ricci-tensor}+(-1)^{P(M+1)}\,\big(&-K^{\,P}_{\phantom{P}MP;N}+(-1)^{P(M+Q)}K^{\,P}_{\phantom{P}QP}\,K^{\,Q}_{\phantom{Q}MN}\\&+(-1)^{NP}K^{\,P}_{\phantom{P}MN;P}-(-1)^{N(M+P+Q)}K^{\,P}_{\phantom{P}QN}\,K^{\,\,Q}_{\phantom{Q}MP}\,\big)\,.
\ea
\ee
In terms of the super-vielbeins, the torsion tensor is given by 
\be\label{addB}
T^{A}_{~BC} E^{C} \wedge E^{B} = T^{A} \equiv d E^{A} + \Omega^{A}_{~B}\wedge E^{B}\,.
\ee


\section{Supermanifolds on given bosonic manifolds }
\label{sec2}

In the present section we are interested in studying examples of supermanifolds 
which have a given bosonic submanifold. We construct a corresponding supermanifold by adding the anticommuting coordinates and by requiring that 
the supermanifold is super-Ricci flat. 

Before proceeding we would like to give an example of a simple model where 
the procedure for constructing a supermanifold is illustrated. Let 
us consider the superfield $\Phi(x,\theta)$ and we set to zero 
all fermionic components. Then, $\Phi = \sum_{n} \Phi_{n} (\theta^{2})^{n}$. We 
can consider the naive generalization of the Klein-Gordon (KG) equation 
$\p_{m} \p^{m} \phi + \frac{1}{3!} \phi^{3}=0$ to a superfield equation 
\begin{equation}\label{exA}
\eta_{AB}\,\p^{B} \p^{A} \Phi + \frac{1}{3!}\,\Phi^{3}_{0} = 
(\p_{m}\p^{m} + \p_{\mu} \p^{\mu}) \Phi + \frac{1}{3!}\,\Phi^{3}=0
\end{equation}
where the contraction of the spinorial indices is performed with 
an invariant tensor (see the discussion in sec. \ref{sec1} after eq.~(\ref{inverse})). 
To be precise, we assume that the supermanifold is $\mathbb{R}^{(3|2)}$ 
and therefore eq. (\ref{exA}) decomposes into two equations
\begin{equation}\label{exB}
\p_{m}\p^{m} \Phi_{0} + \frac{1}{3!}\, \Phi_{0}^{3} + 2\, \Phi_{2} =0\,, 
~~~~
\p_{m}\p^{m} \Phi_{2} + \frac{1}{2!} \,\Phi_{0}^{2} \Phi_{2} =0\,.
\end{equation}
The first equation can be easily solved in terms of $\Phi_{2}$ and, 
by inserting it into the second equation, we get a four-derivative
differential operator (which in principle has ghosts) and 
the field equation
\be\label{addC}
\p^{4} \Phi_{0} + \frac{1}{3!}\, \p^{2} \Phi^{3}_{0} + \frac{1}{2!}\, 
\Phi^{2}_{0}\, \p^{2} \Phi_{0} + \frac{1}{12}\, \Phi^{5}_{0} =0\,.
\ee
 This 
is a weaker condition on $\Phi_{0}$ than the usual KG 
equation and it must be solved in order that the superfield $\Phi_{0}$ 
can be the first component of the superfield $\Phi$ which solves 
the new KG equation. 
The present example has no physical relevance, but it 
illustrates the problem appearing in the extension of a given 
bosonic metric to a supermetric and the extension of the bosonic equations to super-equations. 
Even in this case, we assume 
that the background has no fermions and we add only 
superfields with bosonic components.  

We have to mention that equations similar to (\ref{exA}) where 
already proposed in several papers in the first years of supersymmetry 
(see for example \cite{Arnowitt:1975xg,Nath:1976ci,Arnowitt:1978jq}). 
However, this construction was abandoned since the resulting theories possess ghosts and higher-spin fields (see for example \cite{VanNieuwenhuizen:1981ae} and the 
references therein). Here, we do not pretend 
to interpret the super-Ricci flatness as field equations with dynamical content,  
but we consider these equations  
a way of recasting the informations on the manifold together the RR fluxes 
in a single mathematical structure. This is very similar to the case of NS-NS fluxes and 
the generalized geometry of \cite{Hitchin:2004ut,Gualtieri:2003dx}. 


\subsection{General Framework}

In the present section, we give the general equations 
of the super-Ricci tensor $R_{MN}$ by expanding the 
supermetric in components. These are very useful in order to single out the structure 
of the equations. Indeed, it can be observed that at each order there is free component of the supermetric that is 
used to solve the corresponding equation. 
First, we show that every equation  
at the level lower than the maximum $\theta$-expansion can be 
easily algebraically solved, then we show that the remaining equations 
are high derivatives and they provide consistency conditions 
for the construction of the corresponding 
supermanifold. As in the above example, 
using the iterative equations, one derives the highest derivative equation.  This 
is rather cumbersome since the iterative procedure tends to explode soon in long 
unmanageable expressions. We first tame the linearized system and then we tackle 
the non-linear one. For the former we are able to provide a complete analysis; for 
the latter only the case $(d|2)$ will be discussed.  	

In order to study the structure of the equation for super Ricci-flat manifold 
it is convenient to consider the bosonic and fermionic components separately.   
For that we display the few terms coming from the superfield 
expansion of the supermetric:
\begin{equation}
\begin{aligned}
&G_{mn}=g_{mn}+\frac{1}{2}\,f_{mn\m\n}\,\theta^\n\theta^\m+\mathcal{O}(\theta^4),\\
&G_{m\n}=t_{m\n\r}\,\theta^{\r}+\frac{1}{6}\,w_{m\n\r\s\tau}\,\theta^\tau\theta^\s\theta^\r+\mathcal{O}(\theta^5),\\
&G_{\m\n}=h_{\m\n}+\frac{1}{2}\,l_{\m\n\r\s}\,\theta^\s\theta^\r+\mathcal{O}(\theta^4)\,,\\
\end{aligned}
\end{equation}
where the superfield $g_{mn}$ has the following 
boundary term
\begin{equation}
G_{mn}|_{\theta=0} = g^{(body)}_{mn}
\end{equation}
with $g^{(body)}_{mn}$ the metric of the bosonic submanifold. 
The 
expansion of the superfield $G_{MN}$ contains both commuting and 
anticommuting components, however since we are interested only 
in supermanifold modeled on bosonic manifold plus some $p$-forms, we 
set all the fermionic component to zero. 

Obviously, the metric is defined up to superdiffeomorphisms; 
they are the gauge symmetries under which the 
Ricci tensor is invariant. A superdiffeomorphism is parametrized by a supervector field $\xi_M = (\xi_m, \xi_\m)$ 
where $\xi_m$ and $\xi_\m$ are superfields. The completion from the usual  
symmetries, namely diffeomorphisms and local supersymmetry, is obtained by 
the so-called gauge-completion (see for a review \cite{VanNieuwenhuizen:1981ae}) 
and the complete answer is given by 
\begin{equation}\label{superdiff}   
\delta G_{MN} = G_{MP}\,\xi^P_{~;N} + (-)^{(P+1)N}\,\xi^P_{~;M}\,G_{PN}\,.
\end{equation}
This symmetry is used to impose suitable gauge choices simplifying the algebraic analysis. We found advantageous 
to fix the gauge as in \cite{Grassi:2004ih,Tsimpis:2004gq}. 

At the lowest order in the  odd variables the equation $R_{MN} =0$ 
reads (in App. \ref{appA} the Ricci tensor $R_{\mu m}$ is given) 
\begin{eqnarray}\label{riA}
R_{mn}&=&R_{mn}^{\scriptscriptstyle(body)}-\frac{1}{2}\,h^{\r\s}(h_{\s\r\,;(mn)}- 
t_{m[\s\r];n}-t_{n[\s\r];m}+f_{mn\s\r})\\
&&-\frac{1}{4}\,h^{\m\n}h_{\n\r,m}h^{\r\s}h_{\s\m,n}  
+h^{\m\n}t_{m(\n\r)}h^{\r\s}t_{n(\s\m)}+\mathcal{O}(\theta^2)=0\,, \nonumber
\end{eqnarray}

\begin{equation}\label{riB}
\begin{aligned}
R_{\m\n}=&-\frac{1}{2}\,g^{mn}(h_{\m\n;mn}-2\,t_{m[\m\n];n}+f_{mn\m\n})-\frac{1}{2}\,h^{\r\s}(l_{\m\n\s\r}+l_{\s\r\m\n}+l_{\m\r\s\n}+l_{\s\n\m\r})\\
&-\frac{1}{4}\,h^{\r\s}g^{mn}[(2\,t_{m[\m\n]}-h_{\m\n,m})(2\,t_{n[\s\r]}-h_{\s\r,n})+(2\,t_{m[\m\r]}-h_{\m\r,m})(2\,t_{n[\s\n]}-h_{\s\n,n})]\\
&+\frac{1}{4}\,h^{\r\s}g^{mn}(2\,t_{m(\m\r)}-h_{\m\r,m})(2\,t_{n(\n\s)}-h_{\n\s,n})+\mathcal{O}(\theta^2)=0\,.
\end{aligned}
\end{equation}

Eq. (\ref{riA}) can be easily solved in terms of $f_{mn \r\s}$. In the 
same way,  (\ref{riB}) is solved by a suitable combination of $l_{\m\n\r\s}$. 
At the next order, a free bosonic field will appear 
from the expansion of the superfields $G_{MN}$ and a new equation fixes it. This iterative procedure 
allows us to solve the complete set of equations, starting from a given bosonic metric $g_{mn}$ (and with a flat supermetric in the 
fermionic sector $g_{[\mu\nu]}$), in terms 
of the components of $G_{MN}$. At the end of the iterative procedure, we find the consistency conditions 
on the bosonic manifold. 
We first 
analyze the linearized equations and we expand around the flat space and around a Ricci flat ($R^{(body)}_{mn}$) bosonic space. It 
turns out that variation of the bosonic Ricci tensor must satisfy a set of differential equations (with a 
high-derivative differential operator) and some algebraic conditions (the space must have vanishing scalar curvature). 
However, some of the constraints are weaker in the non-linear theory. 

Furthermore, one can also impose additional structure in the superspace such as a supercomplex 
structure. Some simplifications are in order in the case of a K\"ahler manifolds. 

On a K\"ahler supermanifold the hermitian metric supertensor comes from  
a K\"ahler superpotential, $\mathcal{K}$\,, by the following formula:
\begin{equation}
G_{M\bar{N}}=\mathcal{K}_{\,,\,M\bar{N}}\,.
\end{equation}
As a consequence the Ricci super-tensor components in holomorphic  
coordinates reduce to the form
\begin{equation}
\begin{aligned}
R_{M\bar{N}}&=-(-1)^{P+Q}G^{\,\bar{P}Q}G_{Q\bar{P},M\bar{N}}+( 
-1)^{P+Q+S+M(P+R)}G^{\,P\bar{Q}}G_{\bar{Q}R,M}G^{\,R\bar{S}}G_{\bar{S}P, 
\bar{N}}\\
&=-(\ln\, \mathrm{sdet}\, G)_{,M\bar{N}}
\end{aligned}
\end{equation}

We can expand again the supermetric in the odd coordinates  
($\theta^{\m}, \theta^{\bar \m}$)
\begin{equation}
\begin{aligned}
&G_{m\bar{n}}=g_{m\bar{n}}+h_{\r\bar{\s},m\bar{n}}\,\theta^{\bar{\s}}\theta 
^\r+
\frac{1}{2}\,f_{m\bar{n}\r\s}\,\theta^\s\theta^\r+\frac{1}{2}\,f_{m\bar{n}\bar{\r}\bar{\s}}\,\theta^{\bar{\s}}\theta^{\bar{\r}}+
\mathcal{O}(\theta^4),\\
&G_{m\bar{\n}}=h_{\,\bar{\n}\r,m}\,\theta^\r+t_{m\bar{\n}\bar{\r}}\,\theta^{\bar{\r}}+\mathcal{O}(\theta^3),\\
&G_{\m\bar{n}}=\,t_{\m\bar{n}\r}\theta^\r+h_{\,\m\bar{\r},\bar{n}}\,\theta^{\bar{\r}}+\mathcal{O}(\theta^3),\\
&G_{\m\bar{\n}}=h_{\m\bar{\n}}+l_{\m\bar{\n}\r\bar{\s}}\,\theta^{\bar{\s}}\theta^\r+
\frac{1}{2}\,l_{\m\bar{\n}\r\s}\,\theta^\s\theta^\r+\frac{1}{2}\,l_{\m\bar{\n}\bar{\r}\bar{\s}}\,\theta^{\bar{\s}}\theta^{\bar{\r}}+
\mathcal{O}(\theta^4),\\
&f_{m\bar{n}\bar{\r}\bar{\s}}=f_{n\bar{m}\s\r}^*\,,\qquad  
t_{\r\bar{m}\s}=t_{m\bar{\r}\bar{\s}}^*\,,\qquad  
l_{\m\bar{\n}\bar{\r}\bar{\s}}=l_{\n\bar{\m}\s\r}^*\,,
\end{aligned}
\end{equation}
and obtain the equivalent of the super Ricci-flatness equation,  
$R_{MN}=0$\,, for $g$, $h$, $l$ and $t$\,. The last line expresses the 
constraints coming from the hermiticity of the K\"ahler potential. 
At the lowest order in the  
odd variables one gets
\begin{equation}\label{RF}
\begin{aligned}
R_{m\bar{n}}=&R_{m\bar{n}}^{\scriptscriptstyle(body)}-h^{\bar{\r}\s}\,h_{\s\bar{\r},m\bar 
{n}}-h^{\m\bar{\n}}\,h_{\bar{\n}\r,m}h^{\r\bar{\s}}h_{\bar{\s}\m,\bar{n}}+
\mathcal{O}(\theta^2)\, =\nonumber \\
=&R_{m\bar{n}}^{\scriptscriptstyle(body)}+(\,\ln(\det h))_{,m\bar{n}}+\mathcal{O}(\theta^2)=0\,,\\
R_{\m\bar{\n}}=&-\,g^{\bar{p}q}h_{\m\bar{\n},q\bar{p}}-h^{\bar{\r}\s}\,l_{\s\bar{\r}\m\bar{\n}}-h^ 
{\r\bar{\s}}h_{\bar{\s}\m,p}\,g^{p\bar{q}}h_{\r\bar{\n},\bar{q}}\,+ 
h^{\r\bar{\s}}\,t_{p\bar{\s}\bar{\n}}\,g^{p\bar{q}}\,t_{\r\bar{q}\m}+\mathcal{O}(\theta^2)=0\,,
\end{aligned}
\end{equation}
where $R_{mn}^{\scriptscriptstyle(body)}$ is the Ricci tensor on the  
body of the Ricci-flat K\"ahler supermanifold (SCY) and  
$h=(h_{\m\bar{\n}})$, from which
\begin{equation}
\frac{R^{\scriptscriptstyle(body)}}{2}=h^{\bar{\r}\s}\,h_{\s\bar{\r},m\bar{n 
}}\,g^{m\bar{n}}+h^{\r\bar{\s}}\,h_{\bar{\s}\m,m}h^{\m\bar{\n}}h_{\bar{\n}\r,\bar{n}}\,
g^{m\bar{n}}=- 
h^{\m\bar{\n}}h^{\r\bar{\s}}\,(l_{\r\bar{\s}\m\bar{\n}}+g^{\bar{q}p}t_{\r\bar{q}\m 
}\,t_{p\bar{\s}\bar{\n}}).
\end{equation}
From this equation, one can easily extract the result of 
\cite{Rocek:2004bi,Rocek:2004ha} by 
setting $\m, \n = 1$. This tells us that if there are no enough anticommuting coordinates 
one cannot embed the bosonic manifold into a super-Ricci flat manifold. In addition, this is also 
a counterexample to support the non-existence of the Yau theorem for super-Calabi-Yau spaces. 

To conclude this section, we have to say that the equations (\ref{riA}--\ref{riB}) 
given in the present form and 
expanded to all orders of the anticommuting variables are untreatable and therefore we need 
to use a different technique. In the  appendix we show how to solve the equations using the vielbein-spin connection formalism.


\subsection{Linearized equations}
\label{linearized}
We start from the linear equations from a flat bosonic manifold and we add the fermionic coordinates perturbatively. 
This 
can be done by linearizing the Ricci-flatness condition and solving the corresponding equations. 

As already suggested, all equations can be easily solved except the very last ones and those 
give us some conditions on the bosonic metric. Afterwards, we generalize the construction 
by starting from a Ricci flat bosonic manifold. This example is rather important since allows us 
to study the deformation from a CY space to a super-CY space. 

In order to discuss linearized equations, we suppose to apply a perturbation
\begin{equation}
\delta G_{MN}=H_{MN}
\end{equation}
to a supermetric and study the corresponding variation of the associated Ricci supertensor,
\begin{equation}
\begin{aligned}
\delta R_{MN}=(-1)^{Q+P(M+N+1)}\,\frac{G^{PQ}}{2}\,\big[\,&(-1)^{MN}H_{QN;MP}-(-1)^{Q(M+N)}H_{MN;QP}\\&+H_{QM;NP}-(-1)^{P(M+N)}H_{QP;MN}\,\big]\,,
\end{aligned}
\end{equation}
which, making parity of indices explicit, splits into
\begin{subequations}\label{li3}
\begin{equation}
\delta R_{\,mn}=\frac{1}{2}\,(-1)^{P+Q}\,G^{PQ}\,\big[\,H_{Qm;nP}+H_{Qn;mP}-H_{mn;QP}-H_{QP;mn}\,\big]\,,
\end{equation}
\begin{equation}
\delta R_{\,m\mu}=\frac{1}{2}\,(-1)^Q\,G^{PQ}\,\big[\,H_{Qm;\mu P}+H_{Q\mu;mP}-(-1)^QH_{m\mu;QP}-(-1)^PH_{QP;m\mu}\,\big]\,,
\end{equation}
\begin{equation}
\delta R_{\,\mu\nu}=\frac{1}{2}\,(-1)^{P+Q}\,G^{PQ}\,\big[\,H_{Q\mu;\nu P}-H_{Q\nu;\mu P}-H_{\mu\nu;QP}-H_{QP;\mu\nu}\,\big]\,.
\end{equation}
\end{subequations}
We consider deformations of a super Ricci-flat background realized by adding $d_F$ flat fermionic directions to a Ricci-flat body: in this case \eqref{li3}'s reduce to
\begin{subequations}\label{li4}
\begin{equation}
\begin{aligned}
\delta R_{\,mn}=\,&\,\frac{1}{2}\,\,g^{\,pq}\,\big[\,H_{qm;np}+H_{qn;mp}-H_{mn;qp}-H_{qp;mn}\,\big]\\
&+\frac{1}{2}\,\,j^{\,\rho\sigma}\,\big[\,H_{\sigma m;n\rho}+H_{\sigma n;m\rho}-H_{mn;\sigma\rho}-H_{\sigma\rho;mn}\,\big]\,,
\end{aligned}
\end{equation}
\begin{equation}
\begin{aligned}
\delta R_{\,m\mu}=\,&\,\frac{1}{2}\,\,g^{\,pq}\,\big[\,H_{qm;\mu p}+H_{q\mu ;mp}-H_{m\mu;qp}-H_{qp;m\mu}\,\big]\\
&-\frac{1}{2}\,\,j^{\,\rho\sigma}\,\big[\,H_{\sigma m;\mu\rho}+H_{\sigma\mu;m\rho}+H_{m\mu;\sigma\rho}+H_{\sigma\rho;m\mu}\,\big]\,,
\end{aligned}
\end{equation}
\begin{equation}
\begin{aligned}
\delta R_{\,\mu\nu}=\,&\,\frac{1}{2}\,\,g^{\,pq}\,\big[\,H_{q\mu;\nu p}-H_{q\nu;\mu p}-H_{\mu\nu;qp}-H_{qp;\mu\nu}\,\big]\\
&+\frac{1}{2}\,\,j^{\,\rho\sigma}\,\big[\,H_{\sigma \mu;\nu\rho}-H_{\sigma\nu;\mu\rho}-H_{\mu\nu;\sigma\rho}-H_{\sigma\rho;\mu\nu}\,\big]\,,
\end{aligned}
\end{equation}
\end{subequations}
where $g$ is the metric on the body and $j$ is the flat fermionic supermetric given by
\begin{equation}\label{oddmetric}
j=\text{diag}[\,\underbrace{\sigma_2\,, \dots ,\,\sigma_2\,}_{d_F/2}\,]\,.
\end{equation}
We can now $\theta$-expand \eqref{li4}'s and get
\begin{subequations}\label{li6}
\begin{equation}
\begin{aligned}
\delta R_{\,mn[\tau_1\dots \tau_{2k}]}^{\,(2k)}=\,&\,\frac{1}{2}\,\,g^{\,pq}\,\big[\,H^{\,(2k)}_{qm[\tau_1\dots \tau_{2k}];np}+H^{\,(2k)}_{qn[\tau_1\dots \tau_{2k}];mp}\\&\qquad\quad-H^{\,(2k)}_{mn[\tau_1\dots \tau_{2k}];qp}-H^{\,(2k)}_{qp[\tau_1\dots \tau_{2k}];mn}\,\big]\\
&+\frac{1}{2}\,\,j^{\,\rho\sigma}\,\big[\,(2k+1)\,H^{\,(2k+1)}_{\sigma m[\rho\tau_1\dots\tau_{2k}];n}+(2k+1)\,H^{\,(2k+1)}_{\sigma n[\rho\tau_1\dots\tau_{2k}];m}\\&\qquad\qquad\,\,-(2k+2)(2k+1)\,H^{\,(2k+2)}_{mn[\sigma\rho\tau_1\dots\tau_{2k}]}-H^{\,(2k)}_{\sigma\rho[\tau_1\dots\tau_{2k}];mn}\,\big]\,,
\end{aligned}
\end{equation}
\begin{equation}
\begin{aligned}
\delta R^{\,(2k+1)}_{\,m\mu[\tau_1\dots\tau_{2k+1}]}=\,&\,\frac{1}{2}\,\,g^{\,pq}\,\big[\,(2k+2)\,H^{\,(2k+2)}_{qm[\mu\tau_1\dots\tau_{2k+1}];p}+H^{\,(2k+1)}_{q\mu[\tau_1\dots\tau_{2k+1}] ;mp}\\&\qquad\quad-H^{\,(2k+1)}_{m\mu[\tau_1\dots\tau_{2k+1}];qp}-(2k+2)\,H^{\,(2k+2)}_{qp[\mu\tau_1\dots\tau_{2k+1}];m}\,\big]\\
&-(k+1)\,j^{\,\rho\sigma}\,\big[\,(2k+3)\,H^{\,(2k+3)}_{\sigma m[\mu\rho\tau_1\dots\tau_{2k+1}]}+H^{\,(2k+2)}_{\sigma\mu[\rho\tau_1\dots\tau_{2k+1}];m}\\&\qquad\qquad\,\,+(2k+3)\,H^{\,(2k+3)}_{m\mu[\sigma\rho\tau_1\dots\tau_{2k+1}]}+H^{\,(2k+2)}_{\sigma\rho[\mu\tau_1\dots\tau_{2k+1}];m}\,\big]\,,
\end{aligned}
\end{equation}
\begin{equation}
\begin{aligned}
\delta R_{\,\mu\nu[\tau_1\dots \tau_{2k}]}^{\,(2k)}=\,&\,\frac{1}{2}\,\,g^{\,pq}\,\big[\,(2k+1)\,H^{\,(2k+1)}_{q\mu[\nu\tau_1\dots\tau_{2k}];p}-(2k+1)\,H^{\,(2k+1)}_{q\nu[\mu\tau_1\dots\tau_{2k}];p}\\&\qquad\quad-H^{\,(2k)}_{\mu\nu[\tau_1\dots\tau_{2k}];qp}-(2k+2)(2k+1)\,H^{\,(2k+2)}_{qp[\mu\nu\tau_1\dots\tau_{2k}]}\,\big]\\
&+(2k+1)(k+1)\,j^{\,\rho\sigma}\,\big[\,H^{\,(2k+2)}_{\sigma \mu[\nu\rho\tau_1\dots\tau_{2k}]}-H^{\,(2k+2)}_{\sigma\nu[\mu\rho\tau_1\dots\tau_{2k}]}\\&\qquad\qquad\qquad\qquad\qquad-H^{\,(2k+2)}_{\mu\nu[\sigma\rho\tau_1\dots\tau_{2k}]}-H^{\,(2k+2)}_{\sigma\rho[\mu\nu\tau_1\dots\tau_{2k}]}\,\big]\,.
\end{aligned}
\end{equation}
\end{subequations}
Analizing the recursive structure of \eqref{li6}'s, one realizes that the only independent fields which effectively play a role in constraining $g_{mn}$ are obtained by contracting odd indices with $j$ as follows:
\begin{subequations}\label{li7}
\begin{equation}
S_{mn}^{\,(2k)}=H_{mn[\tau_1\dots\tau_{2k}]}^{\,(2k)}\,j^{\,\tau_1\tau_2}\,\cdots\,j^{\,\tau_{2k-1}\tau_{2k}}\,,
\end{equation}
\begin{equation}
T_{m}^{\,(2k+1)}=H_{m\tau_1[\tau_2\dots\tau_{2k+2}]}^{\,(2k+1)}\,j^{\,\tau_1\tau_2}\,\cdots\,j^{\,\tau_{2k+1}\tau_{2k+2}}\,,
\end{equation}
\begin{equation}
U^{\,(2k)}=H_{\tau_1\tau_2[\tau_3\dots\tau_{2k+2}]}^{\,(2k)}\,j^{\,\tau_1\tau_2}j^{\,\tau_3\tau_4}\,\cdots\,j^{\,\tau_{2k+1}\tau_{2k+2}}\,,
\end{equation}
\begin{equation}
V^{\,(2k)}=H_{\tau_1\tau_2[\tau_3\dots\tau_{2k+2}]}^{\,(2k)}\,j^{\,\tau_1\tau_3}j^{\,\tau_2\tau_4}\,\cdots\,j^{\,\tau_{2k+1}\tau_{2k+2}}\,.
\end{equation}
\end{subequations}
From these definitions one has a few identities:
\begin{equation}\label{li148}
V^{\,(0)}=0\,,
\quad\quad U^{\,(d_F)}=d_F\,V^{\,(d_F)}\,,
\quad\quad S_{mn}^{\,(0)}=\delta g_{\,mn}\,.
\end{equation}
The calculation is simplified by a gauge-fixing of the supermetric (\ref{superdiff}): 
a convenient choice is
\begin{equation}\label{gaugefix}
G_{m\mu}\,\theta^{\mu}=0\,, \quad\quad\quad
G_{\mu\nu}\,\theta^{\nu}\theta^{\mu}=j_{\mu\nu}\,\theta^{\nu}\theta^{\mu}\,,
\end{equation}
which, in terms of variations, becomes
\begin{equation}
H_{m\mu}\,\theta^{\mu}=0\,, \quad\quad\quad
H_{\mu\nu}\,\theta^{\nu}\theta^{\mu}=0\,.
\end{equation}
We notice that some of the fields defined in \eqref{li7} are vanishing because of the gauge-fixing:
\begin{equation}
T_m^{\,(1)}=0\,, \quad\quad
\label{li1413}
U^{\,(0)}=0\,.
\end{equation}
Introducing $S$, $T$, $U$ and $V$ into \eqref{li6}'s, one obtains in principle four field equations but one of them follows from the other ones when Ricci curvature of the body vanishes. The remaining three field equations corresponding to super Ricci-flatness are
\begin{subequations}
\begin{equation}\label{li141}
\begin{aligned}
(2k+2)(2k+1)\,S_{mn}^{\,(2k+2)}=&\,(2k+1)\,\left(\,T_{m;n}^{\,(2k+1)}+T_{n;m}^{\,(2k+1)}\right)-U^{\,(2k)}_{;mn}\\&-g^{\,pq}\left(S_{qm;np}^{\,(2k)}+S_{qn;mp}^{\,(2k)}-S_{mn;qp}^{\,(2k)}-S_{qp;mn}^{\,(2k)}\right),
\end{aligned}
\end{equation}
\begin{equation}\label{li142}
(2k+2)(2k+1)\,\left(U^{\,(2k+2)}+V^{\,(2k+2)}\right)=\Box\left(U^{\,(2k)}+V^{\,(2k)}\right),
\end{equation}
\begin{equation}\label{li143}
\Box\,S^{(2k)~m}_{~~~m}-S^{(2k)~mn}_{~mn;}=\Box\left(U^{\,(2k)}+V^{\,(2k)}\right).
\end{equation}
\end{subequations}
Recalling the identity \eqref{li148} and the gauge-fixing \eqref{li1413}, from here we find
\begin{equation}\label{li1415}
U^{\,(2k)}+V^{\,(2k)}=0\,,\qquad U^{\,(d_F)}=V^{\,(d_F)}=0\,,
\end{equation}
so that \eqref{li142} is automatically solved and \eqref{li143} with $k=0$ shows that, at linear order, the curvature scalar on the body has to be zero:
\begin{equation}\label{li1416}
R^{\,\scriptscriptstyle{(body)}}=0\,.
\end{equation}
Besides this algebraic constraint on the Ricci curvature, also some differential consistency condition has to be taken into account that comes from the remaining constraints. Introducing
\begin{equation}
\hat{S}_{mn}^{\,(2k+2)}\stackrel{\scriptscriptstyle \mathrm{def}}{=}(2k+2)!\,\,S_{mn}^{\,(2k+2)}-(2k+1)!\left(\,T_{m;n}^{\,(2k+1)}+T_{n;m}^{\,(2k+1)}\right)+(2k)!\,\,U^{\,(2k)}_{;mn}
\end{equation}
into \eqref{li141} and \eqref{li143}, one can rewrite them as
\begin{equation}\label{li1418}
\hat{S}_{mn}^{\,(2k+2)}=\Box\,\hat{S}_{mn}^{\,(2k)}+2\,\hat{S}^{\,\,pq}_{(2k)}\,R^{\,\scriptscriptstyle (body)}_{pmqn}\qquad\quad (k>0)
\end{equation}
with 
\begin{equation}\label{li1410}
\hat{S}_{mn}^{\,(2)}=-\,2\,\,\delta R_{\,mn}^{\,\scriptscriptstyle (body)}\,,\qquad \,\hat{S}_{mn}^{\,(d_F+\,2)}=0\,.
\end{equation}
In other words, at linear order, the only condition to require in addition to \eqref{li1416} is
\begin{equation}\label{li1419}
\mathcal{L}_{\scriptscriptstyle R}^{\,d_F/2}\,R^{\,\scriptscriptstyle{(body)}}_{\,mn}=0
\end{equation}
where $\mathcal{L}_{\scriptscriptstyle R}$ is the Lichnerowicz operator defined on Ricci-flat spaces by
\begin{equation}
\mathcal{L}_{\scriptscriptstyle R}\,X_{\,mn}=\Box\,X_{\,mn}+2\,\,X^{\,pq}\,R^{\,\scriptscriptstyle (body)}_{pmqn}\,.
\end{equation}
In conclusion, at linear order, a Ricci-flat manifold can be deformed in the body of a Ricci-flat supermanifold only if conditions \eqref{li1416} and \eqref{li1419} are satisfied, and, in that case, an overlying supergeometrical structure is constrained by \eqref{li1415}, \eqref{li1418} and \eqref{li1410}. As a consequence, if a consistent deformation is possible, an infinite number of Ricci-flat supermanifolds with gauge-fixed supermetric can be constructed over the same 
bosonic body.  

It would be interesting to compare the present result with the 
results of paper \cite{Lindstrom:2005uh}. We plan to explore deeply the 
relation between the super-Ricci flatness equations and the moduli 
space of the supervarieties. 


\subsection{Non-linear equations}

Since a complete general analysis is rather cumbersome, we give the complete expressions only in the case of $(d|2)$-dimensional supermanifolds. We show that equations at the lowest order can be indeed solved by some free components of the superfields and we 
derive the equations for the highest components. 
Following the scheme given in section \ref{linearized}, by substituting the lower-order results 
one gets the final constraints on the metric of the bosonic submanifold.

As in the present case there are just two fermionic directions, the $\t$-expansion of the supermetric with the gauge-fixing \eqref{gaugefix} is given by
\be
\ba
&G_{mn}=g_{mn}+\frac{1}{2}\,G^{(2)}_{mn}\,\t^2,\\
&G_{m\n}=i\,\,G^{(1)}_{m(\n\r)}\,\t^\r,\\
&G_{\m\n}=i\,\,\vare_{\m\n}\left(1+\frac{1}{2}\,f_{(2)}\,\t^2\right)
\ea
\ee
where $\t^2=i\,\,\vare_{\m\n}\,\t^\n\t^\m$. The metric satisfies the gauge fixing condition 
$\theta^\m G_{\m m} =0$ and $G_{\m\n} \theta^\nu = i \e_{\m\n} \theta^\nu$.

In order to make the notation more convenient, 
we introduce the matrix field $A_m$ defined by
\be
(A_m)^\m_{~\n}=\vare^{\m\l}\,G^{(1)}_{m(\l\n)}\,.
\ee
Since $(A_{m})^{\mu}_{~\nu}$ are real $2 \times 2$ traceless matrices they 
form a connection for $\text{SL}(2, \mathbb{R})$ which is the isometry group of the fermionic sector. 
The $0$-order boson-boson and fermion-fermion blocks of the Ricci supertensor are
\be
R_{mn}^{(0)}=R_{mn}+G^{(2)}_{mn}+\tr(A_mA_n)\,,
\ee
\be
R_{\m\n}^{(0)}=i\,\,\vare_{\m\n}\left(\frac{1}{2}\,R-3\,f_{(2)}\right),
\ee
therefore, the super Ricci-flatness condition fixes all $\t^2\!$-order components in terms of the metric $g_{mn}$ and of 
the field $A_m$
\be
G^{(2)}_{mn}=-R_{mn}-\tr(A_mA_n)\,,
\ee
\be
f_{(2)}=-\frac{R}{6}\,.
\ee
As a result any super Ricci-flat gauge-fixed supermetric will have the form
\be
\ba
&G_{mn}=g_{mn}-\frac{1}{2}\,\left(R_{mn}+\tr(A_mA_n)\right)\,\t^2,\\
&G_{m\n}=i\,\,\vare_{\n\l}(A_m)^\l_{~\r}\,\t^\r,\\
&G_{\m\n}=i\,\,\vare_{\m\n}\left(1-\frac{1}{12}\,R\,\t^2\right).
\ea
\ee
Using these relations, after tedious, 
but straightforward calculations one finds some very compact formulae for the high-order components of the Ricci supercurvature:
\be
\ba
R_{mn}^{(2)}=&\,\,\frac{1}{2}\,\Box R_{mn}+R^{\,pq}R_{mpnq}-\frac{1}{3}\,R\,R_{mn}-\frac{1}{6}\,\nabla_m\nabla_n\,R\\
&+\frac{1}{2}\,\tr(F_{mp}F_n^{~p})-\frac{1}{2}\,\tr(A_m\De^pF_{np})-\frac{1}{2}\,\tr(A_n\De^pF_{mp})\,,
\ea
\ee
\be
R^{(1)}_{m\m\n}=\frac{i}{2}\,\,\vare_{\m\l}\,(\De^n F_{mn})^\l_{~\n}\,,
\ee
\be\label{R2FF}
R_{\m\n}^{(2)}=\frac{i}{4}\,\vare_{\m\n}\Big(g^{mn} R^{(2)}_{mn}+\tr(A^m\De^nF_{mn})\Big).
\ee
By $\De_m$ and $F_{mn}$ we simply mean the gauge-covariant derivative and field strength used in non-abelian gauge theories:
\be
\De_m=\nabla_m+[A_m,\,\cdot\,\, ]\,, 
~~~~~~
F_{mn}=\nabla_mA_n-\nabla_nA_m+[A_m,A_n]\,.
\ee
where $[\,\cdot\,,\,\cdot\,]$ is the $SL(2,\mathbb{R})$ Lie product. 
As one can easily see from \eqref{R2FF}, $R_{\m\n}^{(2)}=0$ follows from $R_{mn}^{(2)}=R^{(1)}_{m\n\l}=0$ hence the complete super Ricci-flatness condition is equivalent to the following couple of constraints:
\be
\frac{1}{2}\,\Box R_{mn}+R^{\,pq}R_{mpnq}-\frac{1}{3}\,R\,R_{mn}-\frac{1}{6}\,\nabla_m\nabla_n\,R
=-\frac{1}{2}\,\tr(F_{mp}F_n^{~p})\,,
\ee
\be
\De^nF_{mn}=0\,.
\ee
Thus, when a given manifold does admit a Ricci-flat super-extension, this is not unique even after the gauge-fixing of the supermetric and the corresponding moduli are encoded into a $\text{SL}(2,\mathbb{R})$ gauge theory living on the body. To get the most general Ricci-flat supermetric we perform super-diffeomorphisms along the gauge-fixed directions:
\be
\ba
&G_{mn}=g_{mn}-\frac{1}{2}\,\left[e^{2\phi}\left(R_{mn}+\tr(A_mA_n)-2\,\nabla_m\phi\nabla_n\phi\right)-\nabla_mB_n-\nabla_nB_m\,\right]\t^2,\\
&G_{m\n}=i\left[\,\vare_{\n\l}\,e^{2\phi}\,(A_m)^\l_{~\r}+\vare_{\n\r}(B_m+e^{2\phi}\,\nabla_m\phi)\,\right]\t^\r,\\
&G_{\m\n}=i\,\,\vare_{\m\n}\left[\,e^{2\phi}\left(1-\frac{R}{12}\,e^{2\phi}\,\t^2\right)-\frac{1}{2}\,B_pB^p\,\t^2\right].
\ea
\ee

\subsection{Einstein Manifolds}

We now consider the class of Einstein manifolds,
defined by the condition
\be
R_{mn}=\L\,g_{mn}\,.
\ee
Examples of Einstein spaces 
are the group manifolds and coset manifolds. For them, the Killing-Cartan metric $g_{mn}$ 
is proportional to quadratic combinations of structure constants: 
$g_{mn} \sim f_{mq}^{~~p} f_{np}^{~~q}$. We will start from well-known 
coset manifolds to give some interesting examples. 

For any Einstein manifold, super Ricci-flatness reads
\be\label{sRicciflatEins}
\L^2\left(\frac{d}{3}-1\right)\,g_{mn}=\frac{1}{2}\,\tr(F_{mp}F_n^{~p})\,,
\ee
\be
\De^nF_{mn}=0
\ee
and Ricci-flat supermetrics have the form
\begin{eqnarray}
&G_{mn}=g_{mn}\left(1-\frac{\L}{2}\,e^{2\phi}\,\t^2\right)-\frac{1}{2}\,\left[e^{2\phi}\left(\tr(A_mA_n)-2\,\nabla_m\phi\nabla_n\phi\right)-\nabla_mB_n-\nabla_nB_m\,\right]\t^2,\nonumber\\
&G_{m\n}=i\left[\,\vare_{\n\l}\,e^{2\phi}\,(A_m)^\l_{~\r}+\vare_{\n\r}(B_m+e^{2\phi}\,\nabla_m\phi)\,\right]\t^\r,\nonumber\\
&G_{\m\n}=i\,\,\vare_{\m\n}\left[\,e^{2\phi}\left(1-\frac{d}{12}\,\L\,e^{2\phi}\,\t^2\right)-\frac{1}{2}\,B_pB^p\,\t^2\right].
\end{eqnarray}
When $d>3$, the gauge field $A_m$ has to be turned on in order to fulfill the super Ricci-flatness condition \eqref{sRicciflatEins}, while a trivial solution with $A_m=B_m=\phi=0$ always exists if the bosonic submanifold is 3-dimensional:
\be\label{trivial}
\ba
&G_{mn}=g_{mn}\left(1-\frac{1}{2}\,\L\,\t^2\right),\\
&G_{m\n}=0\,,\\
&G_{\m\n}=i\,\,\vare_{\m\n}\left(1-\frac{1}{4}\,\L\,\t^2\right)\,.
\ea
\ee
This shows a remarkable fact: on one side Einstein 3-folds are never Ricci-flat but on the other side they are always super Ricci-flat, in the sense that they admit Ricci-flat super-extensions. Moreover, formula \eqref{trivial} can be generalized to Einstein manifolds of any odd dimension $d_B=2k+1$ by
\be
\ba
&G_{mn}=\frac{g_{mn}}{1+\frac{\L}{2k}\,\t^2},\\
&G_{m\n}=0\,,\\
&G_{\m\n}=\frac{j_{\m\n}}{1+\frac{\L}{2k}\,\t^2}+\frac{\frac{\L}{2k}\,\t_\m\t_\n}{\left(1+\frac{\L}{2k}\,\t^2\right)^{\!2}}\,,
\ea
\ee
where $d_F=2k$, $j_{\m\n}$ is defined by \eqref{oddmetric}, and $\t_\m=-\,i\,j_{\m\n}\,\t^\n$.


\section{Supermanifolds on Bosonic Coset Manifolds}
\label{sec3}

In the previous sections, we have explored the general case (at the linear level) and 
the $(d|2)$ case at the non-linear level. We expect that the results can be extended to 
the generic case $(d_{B}|d_{F})$. Nevertheless, we would like to present here some applications 
and some examples to support our analysis. The best examples are for instance the supergroup 
manifolds. Since the well-known fact that they are Einstein spaces can be straightforwardly generalized 
to super-Einstein spaces. Furthermore, the proportionality constant between the super-Ricci tensor 
and the metric is proportional to the virtual dimensions of the manifold. In the case 
of odd dimensional real manifold the necessary (but not sufficient) condition  to have vanishing 
Ricci tensor is $d_{B} = d_{F} +1$ where $d_{B}$ and $d_{F}$ are the bosonic and the fermions 
real dimensions. For even dimensional manifold the relation becomes $d_{B} = d_{F} -2$ (see 
\cite{DeWitt:1992cy}). 

Let consider, for example, the spheres $S^{n} = SO(n+1)/SO(n)$, 
the Stiefel manifolds\footnote{They are 
a special case of $V^{(p,q)} = SO(p)/ SO(q-p)$ and also for that we have found the 
supermanifold generalization.} $T^{n} = SO(n+1)/SO(n-1)$, the projective spaces 
$\mathbb{CP}^{n} = SU(n+1)/U(n)$, the space $T^{(p,q)} = SU(2) \times SU(2)/U(1)$ where the charge of subgroup is identified with a combination $p J_{1,3}+ 
q J_{2,3}$ of the  $U(1) \times U(1)$ charges of the subgroup. Then, there is a natural extension of these spaces in the supercoset manifolds 
\begin{eqnarray}\label{scA}
&&\mathbb{S}^{(2n|2n-1)} = \frac{Osp(2n +2| 2n)}{Osp(2n+1| 2n)}\,, ~~~~~
\nonumber \\ 
&&\mathbb{T}^{(2n|2n-1)} =  \frac{Osp(2n +2| 2n)}{Osp(2n | 2n)}\,, ~~~~~
\nonumber \\ 
&&\mathbb{CP}^{(n|n+1)} = \frac{PSU(n+1|n+1)}{SU(n|n+1)}\,.
\end{eqnarray} 
The case of $T^{(1,1)}$ is a case of the second series, where 
$n=1$ and the $SO(2)$ is embedded in the $SO(4)$ as described above. 
Notice that in every case the supergroup in the numerator is super Ricci 
flat (in particular the case of $\mathbb{CP}$ is a super-Calabi-Yau). The 
dimensions of these space is $(2n+1|2n), (4n+1|4n)$ and $(2n|2n +2)$ respectively. 
The dimension of the bosonic submanifold is the dimension of the bosonic 
coset space. To check the Ricci-flatness, the best way is too construct the 
supervielbeins $E^{A}$. 

This can be done as follows: 
we divide the generators of the algebra $T^{A}$ into two subsets: 
$T^{a}$ belonging to the coset space  and $T^{a'}$ belonging to the 
subgroup. Then, we define a group element $g(x,\theta)$ 
where $x,\theta$ are coordinates associated to the 
generators $T^{a}$ where we have distinguished the bosonic and 
fermionic indices. The next step is to construct the 
Maurer-Cartan forms from the differentials 
\be\label{addF}
g^{-1} d g = E^{a} T_{a} + H^{a'} T_{a'}\,,
\ee
where $E^{a}$ are the supervielbeins and $H^{a'}$ are 
the so-called $H$-connections. The metric on the group 
manifold is construct in terms of an invariant 
metric $\eta_{AB}$ on the coset space,  
\begin{equation}\label{scB}
ds^{2} = E^{A} \otimes E^{B} \eta_{AB}\,.
\end{equation}
The computation of the super-Ricci tensor is straightforward since 
the Maurer-Cartan equations lead automatically the 
identification of $R_{MN}$ in terms of the structure constant 
of the group. (see \cite{Castellani:1991et,Berkovits:1999zq}). 

Obviously, the super-Ricci flatness 
concerns both real and complex supermanifolds and we will 
present a simple example: the case of superprojective space 
$\mathbb{CP}^{\,1|2}$. 


\subsection{$\mathbb{CP}^{\,(1|2)}$ as a Ricci-flat supermanifold}

This projective supermanifold endowed with the Fubini-Study Ricci-flat  
K\"ahler supermetric, whose K\"ahler superpotential is
\begin{equation}\label{fubinistudy}
\mathcal{K}=\ln(1+z\bar{z}+\vartheta_1\bar{\vartheta}_1+\vartheta_2\bar{ 
\vartheta}_2)\,,
\end{equation}
must obey equations \eqref{RF}. The resulting components of the  
supermetric are given by
\begin{equation}
\begin{aligned}
&G_{z\bar{z}}=\frac{1}{(1+z\bar{z})^2}\left(1-\frac{1- 
z\bar{z}}{1+z\bar{z}}\,(\vartheta_1\bar{\vartheta}_1+\vartheta_2
\bar{\vartheta}_2)+2\,
\frac{1-2z\bar{z}}{(1+z\bar{z})^2}\,\vartheta_1\bar{\vartheta}_1\vartheta_2\bar 
{\vartheta}_2\right),\\
&G_{z\bar{\vartheta}_a}=- 
\frac{\bar{z}\,\vartheta_a}{(1+z\bar{z})^2}\,\left(1 
-2\,\frac{\vartheta_1\bar{\vartheta}_1+\vartheta_2\bar{\vartheta}_2}{1+z 
\bar{z}}\right),\\
&G_{\vartheta_a\bar{z}}=\frac{z\,\bar{\vartheta}_a}{(1+z\bar{z})^2}\,
\left(1 
-2\,\frac{\vartheta_1\bar{\vartheta}_1+\vartheta_2\bar{\vartheta}_2}
{1+z\bar{z}}\right),\\
&G_{\vartheta_a\bar{\vartheta}_b}=- 
\,\frac{\delta_{ab}}{1+z\bar{z}}\left(1- 
\frac{\vartheta_1\bar{\vartheta}_1+\vartheta_2\bar{\vartheta}_2}{1+z\bar 
{z}}+\frac{2\,\vartheta_1\bar{\vartheta}_1\vartheta_2\bar{\vartheta}_2}{ 
(1+z\bar{z})^2}\right)\\&\qquad\qquad  
+\frac{\bar{\vartheta}_a\vartheta_b}{(1+z\bar{z})^2}\,\left(1 
-2\,\frac{\vartheta_1\bar{\vartheta}_1+\vartheta_2\bar{\vartheta}_2}
{1+z \bar{z}}\right),\\
\end{aligned}
\end{equation}
from which one finds
\begin{equation}
\begin{aligned}
&g_{z\bar{z}}=\frac{1}{(1+z\bar{z})^2}\,,\qquad  
g^{\bar{z}z}=(1+z\bar{z})^2\,,\qquad
t_{z\bar{a}\bar{b}}=t_{a\bar{z}b}=0\,,\\
&l_{a\bar{b}c\bar{d}}=\frac{\delta_{a\bar{d}}\,\delta_{c\bar{b}}- 
\delta_{a\bar{b}}\,\delta_{c\bar{d}}}{(1+z\bar{z})^2}\,,
\qquad 
h_{a\bar{b}}=-\,\frac{\delta_{a\bar{b}}}{1+z\bar{z}}\,,\qquad  
h^{\bar{b}a}=(1+z\bar{z})\,\delta^{\,\bar{b}a}\,, \\
&
h_{z\bar a b} = - \frac{\bar z \delta_{\bar a b}}{(1 + z\bar z)^{2}}\,,
\qquad
h_{\bar z a \bar b} = \frac{z \delta_{a \bar b}}{(1 + z\bar z)^{2}}\,,  
\qquad
h_{a\bar b, m \bar n}
  = - \frac{(1 - z\bar z) \delta_{a\bar b}}{(1+ z\bar z)^{3}}\,.
\end{aligned}
\end{equation}
The second equation \eqref{RF} is identically satisfied and the first  
gives the well known result
\begin{equation}
R^{\,\scriptscriptstyle{\mathbf{CP}^1}}_{z\bar{z}}=\frac{2}{(1+z\bar{z}) 
^2}=2\,g_{z\bar{z}}\,.
\end{equation}
We also compute the contributions of the fluxes (the right hand side of  
\eqref{RF}) and
we have
\begin{equation}
h^{\bar c d} h_{d \bar c, m\bar n} + h^{c \bar d} h_{\bar d e, m} h^{e  
\bar f} h_{\bar f c, \bar n} =
(1 + z \bar z) \delta^{a\bar b} \frac{(1 - z\bar z) \delta_{a\bar  
b}}{(1+ z\bar z)^{3}} +
\frac{ z \bar z \delta_{a\bar b} \delta^{a \bar b}}{ (1 + z\bar z)^{4}}  
=  \frac{ 2 }{(1 + z\bar z)^{2}}
\end{equation}
which exactly cancels the contribution of the bosonic curvature showing  
that
the supermanifold is super-Ricci flat.

\def\CP{$\mathbb{CP}^{\,1|2}$}

The construction of the supermetric from a supervielbein 
construction proceeds as follows. 
We consider here the case \CP for simplicity, but it can be adopted also 
for any $n$. The supergroup manifold $PSU(2|2)$ 
is generated by the bosonic generators 
$m^{\a}_{~\b}, \hat{m}^{\a}_{~\b}$ and the fermionic generators 
$q^{\a}_{~\b}, \hat{q}^{\a}_{~\b}$. They 
satisfy the following commutation relations 
\begin{eqnarray}\label{suB}
&&
[m^{\a}_{~\b}, m^{\g}_{~\d} ] = \delta^{\a}_{\d} m^{\g}_{~\b} - 
\delta^{\g}_{\b} m^{\a}_{~\d}\,, ~~~~~
[\hat{m}^{\a}_{~\b}, \hat{m}^{\g}_{~\d} ] = \delta^{\a}_{\d} \hat{m}^{\g}_{~\b} - 
\delta^{\g}_{\b} \hat{m}^{\a}_{~\d}\,, ~~~~~ \\
&&
[m^{\a}_{~\b}, q^{\d}_{\g}] =-\d^{\a}_{\g} q^{\d}_{\b} + \frac{1}{2} \d^{\a}_{~\b} q^{\d}_{\g}\,, ~~~~ 
[\hat{m}^{\a}_{~\b}, q^{\d}_{\g}] = \d^{\d}_{\b} q^{\a}_{\g} - \frac{1}{2} \d^{\a}_{\b} q^{\d}_{\g}, \nonumber \\
&&
[m^{\a}_{~\b}, \hat{q}^{\d}_{\g} ] = \d^{\d}_{\b} \hat{q}^{\a}_{\g} - 
\frac{1}{2} \d^{\a}_{~\b} \hat{q}^{\d}_{\g}\,, ~~~~ 
[\hat{m}^{\a}_{~\b} ,\hat{q}^{\d}_{\g}] = - \d^{\a}_{\g} \hat{q}^{\a}_{\b} + \frac{1}{2} \d^{\a}_{\b} \hat{q}^{\d}_{\g}\,, ~~~~ \nonumber \\
&&
\{q^{\a}_{~\b}, \hat{q}^{\d}_{~\g} \} = a 
(\delta^{\a}_{~\b} m^{\d}_{~\g} + \delta^{\a}_{~\b} \hat{m}^{\d}_{~\g})\,, ~~~~ 
a^{2} =  -1
\end{eqnarray}
and we assume the following Hermitian conjugation rules
\begin{equation}\label{suC}
(m^{\a}_{~\b})^{\dagger} = - m^{\a}_{~\b}\,, ~~~~
(\hat{m}^{\a}_{~\b})^{\dagger} = - \hat{m}^{\a}_{~\b}\,, ~~~~
(q^{\a}_{~\b})^{\dagger} = \e^{\a\g} \hat{q}_{\g}^{\b}\,, ~~~~
(\hat{q}^{\a}_{~\b})^{\dagger} = \e^{\a\g} {q}_{\g}^{\b}\,, 
\end{equation}

It is convenient to introduce new combinations 
\begin{eqnarray}\label{suD}
&& M^{\a}_{\pm,\b} =  m^{\a}_{~\b} \pm \hat{m}^{\a}_{~\b}\,, ~~~~~~~
Q^{\a}_{\pm,\b} = q^{\a}_{~\b} \pm \hat{q}^{\a}_{~\b}\,, 
\end{eqnarray}
and it is easy to verify that we have the schematic commutation 
relations 
\begin{eqnarray}\label{suE}
&&
[M_{+}, M_{+}] \sim M_{+}\,, ~~~~
[M_{+}, M_{-}] \sim M_{-}\,, ~~~~
[M_{-}, M_{-}] \sim M_{+}\,,  \nonumber \\
&&
[Q_{+}, Q_{+}] \sim M_{+}\,, ~~~~
[Q_{+}, Q_{-}] \sim M_{-}\,, ~~~~
[Q_{-}, Q_{-}] \sim m_{+}\,,   \\
&&
[M_{+},Q_{+}] \sim Q_{+}\,, ~~~
[M_{+},Q_{-}] \sim Q_{-}\,, ~~~
[M_{-},Q_{+}] \sim Q_{-}\,, ~~~
[M_{-},Q_{-}] \sim Q_{+}\,. \nonumber
\end{eqnarray}

This means that we can consider the subgroup 
${\cal H} = \{M_{+}, Q_{+}, {\rm tr} \, M_{-} \}$ (where 
${\rm tr} \, M_{-}$ is the element proportional to the unity) 
and therefore the 
coset is parametrized by $\{M_{-} -  {\rm tr} \, M_{-}, Q_{-}\}$ 
and they correspond to the two bosonic directions and to the four  fermionic 
directions of \CP. Finally, we can compute the supervielbeins associated 
to this model. Let us denote by $g(x, \theta)$ 
a coset representative (see 
\cite{Metsaev:1998it}) and then compute the MC forms
\begin{eqnarray}\label{suF}
g^{-1} \p g = L^{\a}_{-\,\b} (M^{\b}_{- \a} - \delta^{\b}_{~\a} {\rm tr} M_{-}) 
+ \Lambda^{\a}_{-\b} Q^{\b}_{- \a} + 
K^{\Sigma} T_{\Sigma}\,,  
\end{eqnarray}
where $L^{\a}_{-\,b}$ is the MC form associated to the bosonic 
generator of the coset, $\Lambda^{\a}_{-\b}$ is the anticommuting 
MC form associated to the fermionic directions. $T_{\Sigma}$ are the 
generators of the subgroup and $K^{\Sigma}$ are the $H$-connections.  
The supervielbeins are constructed by 
\begin{eqnarray}\label{suG}
L^{\a}_{-\,\b} = 
E^{\a}_{-\,\b M} \p Z^{M}\,, ~~~~
\Lambda^{\a}_{-\,\b} = 
\tilde E^{\a}_{-\,b M} \p Z^M\,, ~~~~
\end{eqnarray}
where $Z^{M}= (X^{\mu}_{~\nu}, \theta^{\mu}_{~\nu})$. So, 
the metric and the $B$ field can be written as 
\begin{eqnarray}\label{suH}
&& 
G_{MN} = 
E^{\a}_{-\,\b (M} \e_{\a\a'} \e^{\b\b'} E^{\a'}_{-\, \b' N)} + 
\tilde E^{\a}_{-\,\b (M} \e_{\a\a'} \e^{\b\b'} \tilde E^{\a'}_{-\, \b' N)}\,, ~~~~
\nonumber \\
&& 
B_{MN} = 
E^{\a}_{-\,\b [M} \e_{\a\a'} \e^{\b\b'} E^{\a'}_{-\, \b' N]} + 
\tilde E^{\a}_{-\,\b [M} \e_{\a\a'} \e^{\b\b'} \tilde E^{\a'}_{-\, \b' N]}\,, ~~~~
\end{eqnarray}

The bosonic coordinates $X^{\mu}_{~\nu}$ can be expressed in term 
of a set of complex coordinates $w,\bar w$, and for the fermionic 
coordinates we also introduce two pairs of complex anticommuting 
coordinates $\vartheta_{i}$. 

In the same spirit, we 
can consider the space 
\be\label{seLAA}
\mathbb{CP}^{(n|n+1)} = \frac{PSU(n+1|n+1)}{SU(n|n+1)}
\ee
which is a super-Calabi-Yau. The metric of this space is obtained 
as follows: condider the generators of $PSU(n+1|n+1)$ which 
are $(J^{A}_{~B}, Q^{A}_{~B'}, K^{A'}_{~B'})$ where $J$ and $K$ and 
the generators of the first and the second $SU(n+1)$. The indices $A,B, A', B'$ run over $1, \dots, (n+1)^{2} -1$. The fermions $Q^{A}_{~B'}$ are 
$(n+1)^{2}$ complex generators. In the subgroup there are the generators 
$J^{a}_{~b}, J_{0}, Q^{a}_{~B'}, K^{A'}_{~B}$. We decompose the generators 
of $PSU(n+1|n+1)$ as follows $J^{A}_{~B} = J^{n+1}_{~b}, J^{a}_{~b}, J^{n+1}_{~n+1}$ and $Q^{A}_{~B'} = Q^{n+1}_{~B'}, Q^{a}_{~B'}$. Therefore, 
we take $ J^{n+1}_{~b},  Q^{n+1}_{~B'}$ as the generator of the coset. Then, 
we have the supervielbeins and the H-connections
\be\label{seLBA}
g^{-1} d g = E^{b} J^{n+1}_{~b} + E^{B'} Q^{n+1}_{~B'} + 
\omega^{a}_{~b} J^{b}_{~a} + \omega^{B'}_{~a} Q^{a}_{~B'} + 
\omega^{A'}_{~B'} K^{B'}_{~A'} + \omega_{0} J_{0}\,, 
\ee
The supervielbeins $(E^{b}, E^{B'})$ form a multiplet of the fundamental 
representation of the $SU(n|n+1)$ and therefore the metric is 
given by 
\be\label{seLC}
ds^{2} = E^{a} \otimes \bar E^{b} \eta_{ab} + 
E^{A'} \otimes \bar E^{B'} \eta_{A'B'}\,.
\ee
where $(\eta_{ab}, \eta_{A'B'})$ is the Killing metric preserved by 
the transformations of $SU(n|n+1)$.

\subsection{$\mathbb{S}^{(3|2)}$ as a Ricci-flat supermanifold}

As a second explicit construction of a super Ricci-flat space, we consider 
the supersphere. This is a generalization of the sphere 
as a coset space $\text{SO}(n+1)/\text{SO}(n)$. We define the supersphere as the 
supercoset $\text{Osp}(4|2)/\text{Osp}(3|2)$. 

The generators of the group $\text{Osp}(4|2)$ are $T^{A}_{~B}, Q^{A}_{\a}, T^{\a}_{\b}$ where $T^{A}_{~B},$ generate $\text{SO}(4)$,  
$T^{\a}_{\b}$ generatr $\text{Sp}(2)$ and $Q^{A}_{\a}$ are the fermions. They 
transform as a vector of $\text{SO}(4)$ and a spinor of $\text{Sp}(2)$. We 
decompose the generators in terms of the subgroup generators as follows  
$(T^{\,a}_{~b}, T^{\,4}_{~a}, Q^{\,a}_{~\a}, Q^{\,4}_{~\a}, T^{\,\a}_{~\b}$). The generators 
of the subgroup are $(T^{a}_{~b}, Q^{a}_{~\a}, T^{\a}_{~\b})$. Therefore, 
we introduce the coordinates of the coset $z^{i}, \theta^{\a}$ and 
we define the superline
\begin{equation}\label{newA} 
Z^{2} = \d_{ab}\,x^{a} x^{b}+ i\,\vare_{\a\b}\,\theta^{\a} \theta^{\b}\,.
\end{equation}
It is easy to compute the 
supervielbeins: one gets
\be\label{newB}
\ba
&E^a=\frac{\sqrt{2}}{1+X^2}\,\,\text{d}x^a\\
&E^\a=\frac{i}{(1+X^2)^\frac{1}{2}}\,\,\left(\text{d}\t^\a-\frac{\t^\a x_m}{1+X^2}\,\,\text{d}x^m\right)
\ea
\ee
with $X^2=x^2+\t^2$ and the metric is given by 
\be\label{newC}
ds^{2} = \d_{ab}\,E^{a} \otimes E^{b} + i\,\vare_{\a\b}\,E^{\a} \otimes E^{\b}.
\ee
The various components of the metric are
\be\label{newD}
\ba
&G_{mn}= \frac{2\,\d_{mn}}{(1 + x^{2})^2}\left( 1 -\frac{2\t^2}{1+x^{2}} \right)+\frac{x_mx_n}{(1+x^2)^3}\,\,\t^2\,,\\
&G_{m\n}=-\,\frac{i\,x_m\t_\n}{(1+x^2)^2}\,, \\
&G_{\m\n}=\frac{i\,\vare_{\m\n}}{1 + x^{2}}  \left( 1 -\frac{\t^2}{1+x^{2}} \right)\,,
\ea
\ee
where $x_m=\d_{mn}\,x^n$ and $\t_\m=\vare_{\m\n}\,\t^{\,\n}$.
Notice that in this case the flux of the $p$-form is normalized to unity and the normalization is fixed by the gauge-fixing. 

It is easy to show that 
the space is super Ricci flat by computing the 
connection and the Ricci tensor. This is another interesting example of a 
space which is rendered a super Ricci flat. Notice that the 
supervielbein $(E^{a}, E^{\a})$ is a metapletic representation 
of the subgroup, namely it is a fundamental representation of the 
supergroup $\text{Osp}(3|2)$. The metric (\ref{newC}) is invariant under the 
subgroup action since the flat metric is invariant under the action of the 
supergroup.

\subsection{$\mathbb{T}^{(1,1|4)}$ as a Ricci-flat supermanifold}

Another interesting space is the Sasaki-Einstein space $T^{(1,1)}$. 
This is identified with the coset 
\be\label{seA}
T^{(1,1)} = \frac{ SU(2) \times SU(2)}{U(1)}\,.
\ee
This space is viewed as an $S^{1}$ fibration over $S^{2} \times S^{2}$ and 
this follows from the classification given by Tian and Yau \cite{Tian:1986ie,Tian:1987if}. 
Notice that the five dimensional regular Sasaki-Einstein space are classified 
completely in terms of the K\"alher-Einstein metric on the base. 
One important example is $S^{5}$ whose base is $\mathbb{CP}^{2}$. Notice that using our technique one can construct the corresponding super-Ricci flat 
space and it turns out that the supersphere is given by 
\be\label{seB}
\mathbb{S}^{(5|4)} = \frac{Osp(6|4)}{Osp(5|4)}
\ee
Notice again the supergroup $Osp(6|4)$ is super-Ricci flat and 
the metric is obtained by deriving the supervielbeins $(E^{i}, E^{\a \a'})$
(the index $i= 1,\dots,5$ and $\a, \a'=1,2$ and they are a vectorial representation of $Osp(5|4)$. Indeed, by decomposing a generic 
group element as 
\be\label{seC}
g = 
\left(
\begin{array}{cc} 
\vspace{.2cm}
g^{A}_{~~B} & g^{A}_{~~\b\b'} \\
g^{\a\a'}_{~~B} & g^{\a\a'}_{~~\b\b'}
\end{array}
\right)\,, ~~~~~~~~
g^{A}_{~~B} \in SO(6)\,, ~~~~ g^{\a\a'}_{~~\b\b'} \in Sp(4)\,,
\ee
the trasformed vielbein can be easily computed. Notice that $Sp(4) \simeq
SO(2,3)$ and therefore the spinorial representation of $Sp(4)$ is isomorphic 
the spinorial representation of $SO(2,3)$. This implies that we can 
pick four fermionic coordinates $\theta^{\a\a'}$. 

Following the bosonic construction, we can see the sphere $S^{5}$ as an 
Hopf fibration of $\mathbb{CP}^{2}$, one can show that 
the supersphere is an $S^{1}$ fibration over the supermanifold $\mathbb{CP}^{(2|2)} = SU(3|2) / U(2|2)$. The bosonic part 
of this coset is $SU(3) \times U(2) / (U(2) \times U(2)) = SU(3)/U(2) = \mathbb{CP}^{2}$. In addition, it has $3 \times 2 - 2\times 2= 2$ complex fermions which can be mapped into 4 real fermions. So, we conclude that 
the supersphere $\mathbb{S}^{(5|4)}$, 
which is super-Ricci flat supermanifold, is a super-Sasaki-Einstein over the 
super-K\"ahler-Einstein $\mathbb{CP}^{(2|2}$.  The condition to be 
super-Sasaki-Einstein is equivalent to state that the Hopf fibration 
of its basis is a super-Ricci flat. This resembles the construction 
of the Calabi-Yau cone over a bosonic Sasaki-Einstein where 
adding a non-compact radial direction, the space is Ricci flat. Here, 
the additional fermionic coordinates guarantee the vanishing of the 
super Ricci tensor. An important issue here is the fact that the bosonic 
submanifold is compact and the non-compactness is in the fermionic 
directions. 

Going back to $T^{(1,1)}$. This is a Sasaki-Einstein space and 
the Calabi-Yau constructed on it is the conifold space. Adding 
one addition bosonic coordinates, one can achieve the Ricci-flatness. 
The topology of the space K\"ahler-Einsteins space of its basis
is $\mathbb{CP}^{1} \times \mathbb{CP}^{1}$, and $T^{(1,1)}$ is a fibration 
over it. The superspace $T^{(1,1|4)}$ can be 
also viewed as an $S^{1}$ fibration over a basis. Indeed, one 
can guess that the basis is the generalization of the above 
considerations and 
\be\label{seE}
\mathbb{CP}^{1} \times \mathbb{CP}^{1} \longrightarrow 
\mathbb{CP}^{(1|1)} \times \mathbb{CP}^{(1|1)}.
\ee
and therefore the superspace $T^{(1,1|4)}$ is the symmetric 
fibration over $\mathbb{CP}^{(1|1)} \times \mathbb{CP}^{(1|1)}$ leading to 
\be\label{seEA}
T^{(1,1|4)} = \frac{Osp(4|2)}{Osp(2|2)}\,,
\ee 
which is super-Ricci flat. Notice that one can also consider an Hopf 
fibration over a single $\mathbb{CP}^{(1|1)}$ and this yields 
\be\label{seEB}
T^{(1,0|4)} = \frac{Osp(4|2)}{Osp(3|2)} \times \mathbb{CP}^{(1|1)} 
\ee
which is not super-Ricci flat and we know that all other spaces, except 
$T^{(1,1)}$ do not yield a supersymmetric vacuum. So, we could 
conclude that if the space turns out not to be a super-Ricci flat supermanifold, then 
the compactification is not supersymmetric. However, a complete 
analysis will be presented in a separate publication. 

Notice that again the number of fermions matches correclty, for 
each factor $\mathbb{CP}^{(1|1)} = SU(2|1) / U(1|1)$ is 
$2 \times 1 - 1 \times 1 = 1$. This can be also seen 
from the explicit for of the metric for $T^{(1,1|4)}$ 
\be\label{seF}
ds^{2} = E^{s} \otimes E^{s} + 
\alpha (E^{I} E^{J} \eta_{IJ} + E^{\a} E^{\b} \e_{\a\b}) + 
\b (\hat E^{I} \hat E^{J} \eta_{IJ} + \hat E^{\a} \hat E^{\b} \e_{\a\b}) 
\ee
where $E^{s}, (E^{I}, E^{\a}), (\hat E^{I}, \hat E^{\a})$ are 
respectively a sinlget and two vector representations of $Osp(2|2)$.  
The form of the metric if $E^{s} = d\sigma + A$, where $\sigma$ is 
the coordinate on $S^{1}$ of the fibration and $A$ is the gauge field 
on $S^{1}$\footnote{The gauge field $A$ is 
a superconnection and therefore it has bosonic and fermionc components 
$A = (A_{I} dy^{I} + A_{\a} d\theta^{a})  + (\hat A_{I} d\hat y^{I} + \hat A_{\a} d\hat\theta^{a})$ where $(y^{I}, \hat y^{I}, \theta^{\a}, \hat\theta^{\a})$ are 
coordinate of $\mathbb{CP}^{(1|1)} \otimes \mathbb{CP}^{(1,1)}$. 
In order to determine the complete supercoonection given the 
bosonic counterparts we impose the usual superspace constraints, namely 
all fermionic components of the field strength are set to zero and we 
require that there is no gluino field, or equivalenty that first componet 
of the spinorial field strength is set to zero. 
}, shows that the even the superspace has the same structure of the orginal bosonic space. 

\subsection{M-theory and seven dimensional supermanifolds}
 
Similar considerations apply for the M-theory seven dimensional 
spaces. Let us consider first the super-sevensphere $\mathbb{S}^{(7|6)}$. 
In order that this space is super-Ricci flat we need six fermionic 
coordinates and this can be clearly obtained from the 
supercoset
\be\label{seD}
\mathbb{S}^{(7|6)} = \frac{Osp(8|6)}{Osp(7|6)}\,.
\ee
Notice that the supergroup $Osp(8|4)$ is the supergroup compactification 
of the supermembrane on $AdS_{4} \times S^{7}$. But this is not super-Ricci 
flat, on the other hand $Osp(8|6)$ is super-Ricci flat. Again, it 
can be viewed as an Hopf fibration of a K\"ahler-Einstein space 
$\mathbb{CP}^{(3|3)} = SU(4|3) / U(3|3)$. This space has 
$4 \times 3  - 3\times 3 = 3$ complex fermions which are mapped into 6 real fermions of $\mathbb{S}^{(7|6)}$.

Another notable example is the space 
\be\label{seEAAA}
M^{pqr} = \frac{SU(3) \times SU(2) \times U(1)}{SU(2) \times U(1) \times U(1)}
\ee
which is a real sevem dimensional space. According to the general rule, the 
number of fermions needed to make it super-Ricci flat is six. 
This means that one way to converted the bosonic space 
into a supermanifold we can guess the following form 
\be\label{seFAA}
M^{(pqr|6)} = \frac{SU(3|2)}{SU(1|2) \times SU(1|1)}\,.
\ee
To write the super metric we have to decompose the generators of 
$SU(3|2)$ into those of the subgroup. We have $(T_A, T_a, T_0, Q^I_j)$ 
where $T_A \in su(3)$, $T_a \in su(2)$ (in the adjoint repr.) and $Q^I_j$ are in the 
fundamental representation of $SU(3) \times SU(2)$. In addition, there is a charge 
carried by the fermions and generated by $T_0$. The generators of $su(3)$ are decomposed into $(T_8, T'_A)$ where $T_8$ is one of the generators of the 
Cartan subalgebra of $su(3)$. The fermions are decomposed into $Q^I_j = 
(Q^3_j, Q^i_j - \delta^i_j {\rm tr} Q, {\rm tr} Q)$. Therefore, the generators 
of the subgroup are $(T_8, T_a, Q^3_j) \oplus (T_0, {\rm tr} Q)$. So, the 
supervielbeins are organized as follows
\be\label{seG}
(E^{A'}, E^i_j) = \big(e^s, (e^1, f^1_i), (e^2, f^1_i),(e^3, f^1_i) \Big)
\ee
where $e^\Lambda$ ($\Lambda = 1,2,3$) are complex vielbeins 
corresponding to the generators 
$(T^1 \pm i T^2, T^4 \pm i T^5, T^6 \pm i T^7)$ and $f^\Lambda_i$ are the fermions 
in a fundamental representation of $SU(2)$. In this way we see that the 
the supervielbeins are organized according to the representations of $SU(1|2)$. 
The field vielbein $e^s$ is a singlet. The decomposition (\ref{seG}) is 
aslo a representation of the second subgroup. Therefore, the 
complete supermetric is given by 
\be\label{seH}
ds^2 = e^s \otimes e^s + \sum_\Lambda g_\Lambda
\Big(
e^\Lambda \otimes \bar e^\Lambda + f^\Lambda_i \otimes f^\Lambda_j \e^{ij} \Big) 
\ee
where $g_\Lambda$ have to be tuned in a way such that the metric 
is super-Ricci flat. However, it has to be checked if this is really possible. 

There are other example as
\be\label{seI}
M^{100|6} = \frac{Osp(6|4)}{Osp(5|4)} \times \mathbb{CP}^{(1|1)}\,, 
\quad\quad
M^{010|6} = \frac{Osp(4|2)}{Osp(3|2)} \times \mathbb{CP}^{(2|2)}
\ee
and they have different topologies: the first one has the topology $S^5 \times S^2$, 
and therefore it can be view as a fibration over $\mathbb{CP}^2 
\times \mathbb{CP}^1$, the 
second one has the topology $S^3 \times \mathbb{CP}^2$ and it 
can be viewed as an $S^1$ fibration of $\mathbb{CP}^1 \times \mathbb{CP}^2$. 
The two space constructed above are not super-Ricci flat, but 
they are super-Einstein spaces due to the presence of $\mathbb{CP}^{(n|n)}$. 

Notice that according to the previous considerations,  we have that 
a super-Hopf fibration can be derived as follows
\be\label{seL}
\frac{Osp(2 n + 2| 2n)}{Osp(2 n+ 1|2n)} = (
S^1 \hookrightarrow \mathbb{CP}^{(n|n)} )= (S^1 
\hookrightarrow \frac{SU(n+1|n)}{SU(n|n)}) 
\ee  
The two spaces in (\ref{seI}) are the two extreme case when the fibration 
is either over $\mathbb{CP}^2$ or over $\mathbb{CP}^1$. In the other case, 
one has a mix of the two superspace which however cannot be written in a simple 
way. 

As an example of the super-Hopf let us consider the case of supersphere $n=1$ in the 
above formula which is a super-Hopf fibration of $\mathbb{CP}^{(1|1)}$. In particular, we 
can view the supersphere as the locus in $\mathbb{R}^{(4|2)}$ given by the 
curve\footnote{The coordinates $X_I$ have to be considered as 
even objects and not as numbers. This can be done using the technique of 
supermanifolds with a fix number of odd generators.} 
\be\label{seLA}
\sum_{I =1}^{4} X^2_I + \e_{\a\b} \theta^\a \t^\b = 1
\ee
The supersphere has the isometry $Osp(4|2)$. On the other side, the 
space $\mathbb{CP}^{(1|1)}$ is a super-K\"ahler and it is obtained by 
embedded the curve 
\be\label{seLB}
\sum_{i =1}^{3} X^2_i + \theta \bar\t = 1
\ee
in the superspace $\mathbb{R}^{3|2}$ which has the isometry $Osp(3|2)$. The 
Hopf map is obtained by 
setting the following identification
\be
\ba\label{seLCA}
&x_1 = 2 (X_1 X_2 + X_3 X_4) - (X_1 X_4 - X_3 X_2) \e_{\a\b}\theta^\a  \theta^\b\,, 
\quad\\
&x_2 = 2 (X_1 X_4 - X_3 X_2) + (X_1 X_2 + X_3 X_4) \e_{\a\b}\theta^\a  \theta^\b\,, 
\\
&x_3 = X_1^2 + X_3^2 - X_2^2 - X_4^2\,, 
\\
&\theta = \theta^1 + i \theta^2\,, 
\\ 
&\bar\theta = \theta^1 - i \theta^2\,, 
\ea
\ee
In this way, one can see that there is the Hopf map between the 
$\mathbb{S}^{(3|2)}$ and $\mathbb{CP}^{(1|1)}$. One can also see that 
for each point of $\mathbb{CP}^{(1|1)}$ there is a circle in the bigger space. 
Notice that the circle is purely bosonic. However, one can also find a circle (one 
parameter surface in the bigger space) with fermionic direction. Namely, the 
circle extends also in the fermionic part of the supermanifold. In the 
same way, one can show the Hopf fibrations for the other coset spaces. 

If we consider the space (\ref{seD}) as part of the vacuum of 11d supergravity, 
we have to discuss also the rest of the space. The space 
$\mathbb{S}^{(7|6)}$ has only 6 fermions and therefore, the rest of the 
superspace should have 26 fermions. This, however, is not a multiple 
of 4 and there is no such a supersymmetric vacuum. One possible 
choice is to consider the superspace $\mathbb{CP}^{(3|4)}$ and construct an Hopf fibration on it.
This resembles the construction given in \cite{Nilsson:1984bj}. 
There, the compactification of $\mathbb{CP}^{3}$ is studied. The 
resulting theory is $N=3$ on $AdS_{4}$ with RR fluxes. The natural 
superspace construction is to replace $\mathbb{CP}^{3}$ with 
$\mathbb{CP}^{(3|4)}$ and the $AdS_{4}$ with $Osp(6|4) / SO(6) \times SO(1,3)$. 
The solution of type IIA can be embedded in M-theory by an 
Hopf fibration over $\mathbb{CP}^{3}$ obtaining a compactifcation on a 
$S^{7}$. Notice that the superspace 
\be\label{seLAAA}
\frac{Osp(6|4)}{SO(6) \times SO(1,3)} \times \mathbb{CP}^{(3|4)}\,,
\ee
has already 32 fermions and it is not super-Ricci flat. Indeed, while 
the factor $\mathbb{CP}^{(3|4)}$ is a super-Calabi-Yau, the first 
factor is not super-Ricci flat unless a torsion is added (see \cite{Berkovits:1999zq}). So, it is conceivable that we can construct a new 
supermanifold that compensate the non-Ricci flatness of the 
first factor by adding an additional bosonic coordinate. On the 
other hand, we can render the second factor non-Ricci flat.  

\subsection{F-theory and nine dimensional supermanifolds}

Following the suggestions of the previous sections, we can 
consider the F-theory in 12-dimensions. One can compactify the 
theory on a nine sphere in order to preserve the supersymmetry. 
It is natural to consider the space 
\be\label{fA}
\mathbb{CP}^2 \times \mathbb{CP}^2
\ee 
which is a symmetric product of compact spaces and this resembles the case of 
$T^{(1,1)}$. In particular one can construct a space which is a fibration over 
(\ref{fA})
\be\label{fB}
\frac{SU(3) \times SU(3) \times U(1)}{SU(2) \times SU(2) \times U(1) \times U(1)}
\ee
which is a nine dimensional space. Obviously, there is also the nine sphere $SO(10)/SO(9)$ and this is obtained by an Hopf fibration over $\mathbb{CP}^4$. Therefore, 
in this case we have the following supermanifold
\be\label{fC}
S^9 \hookrightarrow \mathbb{S}^{(9|8)} = \frac{Osp(10|8)}{Osp(9|8)}\,.
\ee 
The space is super Ricci flat and it contains eight fermions. The metric can be computed 
in the usual way (see above). In addition, we can construct a superHoft fibrations 
starting by (\ref{fA}) and we have the two cases 
\be\label{fD}
R^{9|8} = \frac{Osp(6|4)}{Osp(5|4)} \times \mathbb{CP}^{(2|2)}\,, \quad\quad
\hat R^{9|8} = \frac{Osp(6|4)}{Osp(4|4)}
\ee
Notice that the first one is not super-Ricci flat since $\mathbb{CP}^{(2|2)}$ is 
a super-K\"ahler-Einstein space  and the first factor is super-Ricci flat. The 
second one is exactly one of the member of the series $\mathbb{T}^{(2n|2n -1)}$ 
given in (5.58). 


\section{RR fields and supermetric}
\label{sec4}

We have already suggested that the flat supermetric in the 
fermionic direction is not a model-independent concept and 
therefore we would rather not use it. On the other side, for the 
applications we are interested in, we have to our disposal 
the RR field strengths that can indeed provide some 
bispinors. For example, the 
case of IIB supergravity in 10d, we have the following RR 
field content: 
\be\label{BB}
F^{\mu\nu} = \gamma^{\mu\nu}_m F^m + \gamma^{\mu\nu}_{mnp} F^{mnp} + 
\gamma^{\mu\nu}_{[mnpqr]} F^{[mnpqr]}\,,
\ee
where $F^m,  F^{[mnp]}$ and $F^{[mnpqr]}$ are 1-, 3- and 
5-forms. They are the field strength of the RR fields. Notice that 
the Dirac matrices $\gamma^{\mu\nu}_m$ and $\gamma^{\mu\nu}_{[mnpqr]}$ are symmetric 
in the spinorial indices while $\gamma^{\mu\nu}_{mnp}$ is antisymmetric. Therefore, it is natural 
to identify non-anticommutativity of the Grassmann coordinates $\theta^\mu$  with the 
symmetric part of the RR field strength and the first component of the supermetric 
in the fermionic direction with the antisymmetric part of the RR field strength
\be\label{addE}
g_{\mu\nu}(x, \theta) = \g^{[mnp]}_{\mu\nu} F^{[mnp]} + {\cal O}(\theta)\,.  
\ee
We have to notice that by the choice (\ref{addE}) we have to change 
the gauge fixing  into 
\be\label{addG}
g_{\m\n}(x,\theta) \theta^\mu \theta^\nu = \g^{[mnp]}_{\mu\nu} F^{[mnp]} \theta^\mu \theta^\nu\,.
\ee
The consequences of these choices will be presented somewhere else where 
the construction of the supermetric from string theory will be presented. 
It is however very satisfactory to see that all RR fields have an interpretation from 
a supergeometry point of view.


\acknowledgements

We thank P. Aschieri, L. Castellani, G. Dall'Agata, P. Fr\'e, A. Lerda, G. Policastro, Y. Oz 
and E. Scheidegger for several discussions on aspects of the present problem. We 
would like to thank J. Evslin for some suggestion about the super-Hopf 
fibration. In addition, we thank the math/phys group R. Catenacci, M. DeBernardi and 
D. Matessi at University of Piemonte Orientale and U. Bruzzo 
for joint seminars and discussion on supermanifolds and supergeometry.

This work is partially supported by the European Community's Human Potential
Programme under contract MRTN-CT-2004-005104 and
by the Italian MIUR under contract PRIN-2005023102. 

\vfill


\appendix
\section{Equation for $R_{m b}=0$ at the lowest order in $\theta$.}
\label{appA}

In this section, we display the formula for the lowest order exapnsion 
of the mixed fermion-boson component of the super Ricci tensor. Again, from this equation 
we can find that there are enough free parameters that can be fixed to solve it algebraically. 

\begin{eqnarray}
&&R_{mb}=\,\,\Bigg[\,\frac{1}{2}\,g^{\,pq}\,(-\,t_{mbg;qp}+t_{qbg;mp}-f_{qpbg;m}+f_{mpbg;q}) 
\nonumber \\
&&
+\frac{1}{4}\,g^{\,pq}\,h^{ef}(2\,t_{q(eb)}+h_{eb,q})(t_{mfg;p}+t_{pfg;m}-f_{mpfg})\nonumber \\
&&-\frac{1}{4}\,g^{\,pq}\,h^{ef}(2\,t_{m(eb)}+h_{eb,m})(t_{qfg;p}+t_{pfg;q}-f_{qpfg})\nonumber \\
&&-\,\frac{1}{2}\,h^{cd}(w_{mbdcg}+w_{mcdbg}+l_{dcbg,m}+l_{dbcg,m})\nonumber \\
&&-\frac{1}{4}\,g^{\,pq}h^{cd}(2\,t_{q[dc]}-h_{dc,q})(t_{pbg,m}-t_{mbg,p}+f_{pmbg})\nonumber \\
&&-\frac{1}{4}\,g^{\,pq}h^{cd}(2\,t_{q[db]}-h_{db,q})(t_{pcg,m}-t_{mcg,p}+f_{pmcg})\nonumber \\
&&-\frac{1}{4}\,h^{cd}\,h^{\,ef}\,(2\,t_{m(eb)}+h_{eb,m})(l_{fdcg}-l_{fcdg}-l_{dcfg})\nonumber \\
&&+\frac{1}{4}\,h^{cd}\,h^{\,ef}\,g^{\,pq}\,t_{qfg}\,(2\,t_{m(eb)}+h_{eb,m})(2\,t_{p[dc]}-h_{dc,p})\nonumber \\
&&-\frac{1}{4}\,h^{cd}\,h^{\,ef}\,(2\,t_{m(ec)}+h_{ec,m})(l_{fdbg}-l_{fbdg}-l_{dbfg})\\
&&+\frac{1}{4}\,h^{cd}\,h^{\,ef}\,g^{\,pq}\,t_{qfg}\,(2\,t_{m(ec)}+h_{ec,m})(2\,t_{p[db]}-h_{db,p})\nonumber \\
&&+\,\,\frac{1}{2}\,g^{\,pq}\,h^{cd}\,t_{qcg}\,(f_{\,mpdb}+h_{\,db;mp}+2\,t_{mbd;p}-2\,t_{pdb;m}-t_{pbd;m}+t_{mdb;p})\nonumber \\
&&+\frac{1}{4}\,g^{\,pq}\,h^{cd}\,h^{ef}\,t_{qcg}\,(2\,t_{m(ed)}+h_{ed,m})(2\,t_{p(fb)}+h_{fb,p})\nonumber \\
&&-\frac{1}{2}\,g^{\,pq}\,h^{cd}\,h^{ef}\,t_{qcg}\,(2\,t_{m(eb)}+h_{eb,m})(2\,t_{p(fd)}+h_{fd,p})\Bigg]\,\,\theta^{\,g}+\mathcal{O}(\theta^3)=0\,, \nonumber 
\end{eqnarray}


\section{Expansion using supervielbeins}
\label{appB}

In the present appendix, we give an alternative way of proceeding using the 
supervielbeins-superconnection description of Ricci tensor. This technique seems 
to be useful, when the supervielbeins can be easily computed. One part of the system 
is simple to solve, but the complete system has the same problems as the metric approach 
used in the text.  

As in the present case when there are just two fermionic directions, 
the supervielbein and superconnection 
$\t$-expansions are given by
\be\label{expvielbein}
\ba
&E^{\,a}_{~m}=\zero\!E^{\,a}_{~m}+\two\!E^{\,a}_{~m}\,\theta^2\,,\\
&E^{\,a}_{~\m}=\one\!E^{\,a}_{~\m\n}\,\theta^\n\,,\\
&E^{\,\a}_{~m}=\one\!E^{\,\a}_{~m\n}\,\theta^\n\,,\\
&E^{\,\a}_{~\m}=\zero\!E^{\,\a}_{~\m}+\two\!E^{\,\a}_{~\m}\,\theta^2
\ea
\ee
and
\be\label{expspinconn}
\ba
&\O^{\,a}_{~bm}=\zero\O^{\,a}_{~bm}+\two\O^{\,a}_{~bm}\,\theta^2\,,\\
&\O^{\,a}_{~b\m}=\one\O^{\,a}_{~b\m\n}\,\theta^\n\,,\\
&\O^{\,a}_{~\b m}=\one\O^{\,a}_{~\b m\n}\,\theta^\n\,,\\
&\O^{\,a}_{~\b\m}=\zero\O^{\,a}_{~\b\m}+\two\O^{\,a}_{~\b\m}\,\theta^2\,,\\
&\O^{\,\a}_{~bm}=\one\O^{\,\a}_{~bm\n}\,\theta^\n\,,\\
&\O^{\,\a}_{~b\m}=\zero\O^{\,\a}_{~b\m}+\two\O^{\,\a}_{~b\m}\,\theta^2\,,\\
&\O^{\,\a}_{~\b m}=\zero\O^{\,\a}_{~\b m}+\two\O^{\,\a}_{~\b m}\,\theta^2\,,\\
&\O^{\,\a}_{~\b\m}=\one\O^{\,\a}_{~\b\m\n}\,\theta^\n\,.
\ea
\ee
where $\zero\!E^{\,a}_{~m}$ and $\zero\O^{\,a}_{~bm}$ are the ordinary vielbein and spin-connection over the body $\mathcal{M}^{\,m}$\,. In order to simplify calculations, one can fix the gauge with respect to superdiffeomorphisms assuming that
\be
E^A_{~\m}\,\t^\m=\d^A_{~\m}\,\t^\m
\ee
and
\be
\O^A_{~B\m}\,\t^\m=0\,,
\ee
which in terms of components is equivalent to set $\,\one\!E^{\,a}_{~[\m\n]}=0\,$, $\,\zero\!E^{\,\a}_{~\m}=\d^{\,\a}_{~\m}\,$, $\zero\O^{\,a}_{~\b}=\zero\O^{\,\a}_{~b}=0\,$, $\one\O^{\,a}_{~b[\m\n]}=\one\O^{\,\a}_{~\b[\m\n]}=0\,$.
To solve the super Ricci-flatness condition, we have to deal with the definition \eqref{defspinconn} of spin-superconnection and the equation
\be\label{superricciflat}
(-1)^{M(B+1)}\,E^M_{~A}\,\left(\,\O^A_{~B[M,N]}-(-1)^{M(B+C)}\,\O^A_{~C[M}\wedge\O^{\,C}_{~BN]}\,\right)=0
\ee
obtained by contracting a couple of superindices in \eqref{defriemann}. Introducing expansions \eqref{expvielbein} and \eqref{expspinconn} into \eqref{defspinconn} one gets eleven equations one of which is nothing but the usual definition of spin-connection over the body manifold $\mathcal{M}^{\,m}$ while the other ones can be solved algebraically by using $\two\!E^{\,a}_m$\,, $\one\!E^{\,a}_{(\m\n)}$\,, $\one\!E^{\,\a}_{m\n}$\,, $\two\!E^{\,\a}_\m$\,, $\two\O^a_{~bm}$\,,  $\one\O^a_{~\b m\n}$\,, $\zero\O^\a_{~\b m}$\,, $\one\O^a_{~b\m\n}$\,, $\two\O^a_{~\b\m}$\,, $\two\O^\a_{~b\m}$\,. Then we have five more equations coming from $\t$-expansion of \eqref{superricciflat}. The three lowest order ones corresponding to boson-boson, boson-fermion and fermion-fermion sectors of the Ricci supercurvature respectively read
\be
\zero\!R_{\,bn}=-\,\one\O^{\,\a}_{~bn\a}\,, ~~~~~~
\one\O^{\,a}_{~\b a\n}=\one\O^\a_{~\b(\a\n)}\,,
\ee
\be
\nabla_{[a}\,\,\one\O^{\,a}_{~\b n]\r}+\two\O^\m_{~\b n}\,\vare_{\m\r}-\frac{1}{2}\,\nabla_n\,\one\O^{\,\m}_{~\b\m\r}+\frac{1}{2}\,\,\one \!E^{\,\m}_{~a\r}\,\,\one\O^a_{~\b n\m}+\one\!E^{\,m}_{~\a\r}\,\zero\!R^{\,\a}_{~\b mn}=0
\ee
and it is evident that also these equations can be solved algebraically by $\,\one\O^{\,\a}_{~bn\a}$\,, $\one\O^\a_{~\b(\a\n)}$\,, $\two\O^\m_{~\b n}$. In conclusion, the whole system of conditions boils down to a couple of equations corresponding to highest order contributions to the Ricci supertensor: in particular from the boson-boson and fermion-fermion sectors one respectively obtains
\be
\ba
&\nabla_{[a}\two\O^{\,a}_{~bn]}+\one\O^{\,a}_{~\g[a\r}\,\one\O^{\g}_{~bn]\s}\,\vare^{\r\s}+\two\!E_{~a}^{m}\,\zero\!R^{\,a}_{~bmn}-\frac{1}{2}\,\nabla_n\two\O^{\,\a}_{~b\a}\\&+\frac{1}{2}\,\,\one\O^{\,\a}_{~\g\a\r}\,\one\O^{\,\g}_{~bn\s}\,\vare^{\r\s}-\frac{1}{2}\,\,\one\O^\a_{~cn\r}\,\one\O^c_{~b\a\s}\,\vare^{\r\s}+\frac{1}{2}\,\two\!E^{\,\m}_{~\a}\,\,\one\O^\a_{~bn\m}\\
&-\frac{1}{2}\,\,\one\!E^{\,\m}_{~a\m}\,\two\O^a_{~bn}-\frac{1}{2}\,\,\one\!E^{\,\m}_{~a\r}\,\nabla_n\,\one\O^a_{~b\m\s}\,\vare^{\r\s}+\,\one\!E^{\,m}_{~\a\r}\,\nabla_{[m}\,\one\O^\a_{~bn]\s}\,\vare^{\r\s}=0
\ea
\ee
and
\be
\ba
&\frac{1}{2}\,\nabla_a\two\O^a_{~\b\n}+\frac{1}{2}\,\,\one\O^a_{~\g a\r}\,\one\O^\g_{~\b\n\s}\,\vare^{\r\s}-\frac{1}{2}\,\,\one\O^a_{~c \n\r}\,\one\O^c_{~\b a\s}\,\vare^{\r\s}\\&-\frac{1}{2}\,\two\!E^{\,m}_{~a}\,\,\,\one\O^a_{~\b m\n}+\one\O^\a_{~\g(\a}\,\one\O^\g_{~\b\n)}+\two\!E^{\,\m}_{~\a}\,\,\,\one\O^\a_{~\b (\m\n)}\\&-\one\!E^{\,\m}_{~a(\m}\two\O^a_{~\b\n)}+\frac{1}{2}\,\,\one\!E^{\,m}_{~\a\r}\,\nabla_m\,\one\O^\a_{~\b\n\s}+\one\!E^{\,m}_{~\a\n}\,\two\O^\a_{~\b m}=0\,.
\ea
\ee
Such a mechanism works recursively in any fermionic dimension, so in general the consistency condition which determines when an ordinary manifold can be embedded into a super Ricci-flat supermanifold consists also at non-linear level of one tensorial and one scalar differential equation for the Ricci tensor of the body manifold $\mathcal{M}^{\,m}$.

\references

\end{document}